\documentclass[aps,twocolumn,showpacs,preprintnumbers,nofootinbib,float,longbibliography,superscriptaddress]{revtex4-2}

\usepackage{epsfig}
\usepackage{graphicx}
\usepackage{palatino}
\usepackage[english]{babel}
\usepackage{hyphenat}
\usepackage{amsmath}
\usepackage{amssymb}
\usepackage{mathtools}
\usepackage{mathrsfs}
\usepackage{bbm} 
\usepackage{bm}
\usepackage{slashed}
\usepackage{epstopdf}
\usepackage{xcolor}
\usepackage{booktabs}
\usepackage{float}
\definecolor{lcolor}{rgb}{0.,0.0,0.}
\definecolor{citcolor}{rgb}{0,0.,0.5}
\usepackage[breaklinks,colorlinks,urlcolor=blue,citecolor=blue,linkcolor=blue]{hyperref}
\usepackage{multirow}
\usepackage{ltablex}
\usepackage{soul}

\newcommand{\Hcal}{\mathcal{H}}
\newcommand{\Ocal}{\mathcal{O}}

\newcommand{\Fcal}{\mathcal{F}}

\newcommand{\Kcal}{\mathcal{K}}

\newcommand{\vect}[1]{\boldsymbol{#1}_{\perp}}
\newcommand{\kt}{\boldsymbol{k}}
\newcommand{\pqt}{\boldsymbol{k_{q}}}
\newcommand{\pqbt}{\boldsymbol{k_{\bar{q}}}}
\newcommand{\pgt}{\boldsymbol{k_{g}}}

\newcommand{\pgtone}{\boldsymbol{k_{g1}}}

\newcommand{\pgttwo}{\boldsymbol{k_{g2}}}
\newcommand{\Pt}{\boldsymbol{P}}

\newcommand{\lt}{\vect{l}}

\newcommand{\bt}{\boldsymbol{b}}

\newcommand{\Kt}{\boldsymbol{K}}

\newcommand{\xt}{\boldsymbol{x}}

\newcommand{\ptk}{\boldsymbol{p_{k}}}
\newcommand{\ptone}{\boldsymbol{p_{1}}}
\newcommand{\pttwo}{\boldsymbol{p_{2}}}

\newcommand{\phtk}{\boldsymbol{p_{hk}}}
\newcommand{\phtone}{\boldsymbol{p_{h1}}}
\newcommand{\phttwo}{\boldsymbol{p_{h2}}}

\newcommand{\calN}{\mathcal{N}}

\newcommand{\der}{\mathrm{d}}

\newcommand{\Tr}{\mathrm{Tr}}

\allowdisplaybreaks

\begin{document}

\title{Probing gluon saturation with forward di-hadron correlations in proton-nucleus collisions}

\author{Paul Caucal}
\affiliation{SUBATECH UMR 6457 (IMT Atlantique, Universite de Nantes, IN2P3/CNRS), 4 rue Alfred Kastler, 44307 Nantes, France}

\author{Zhong-Bo Kang}
\affiliation{Department of Physics and Astronomy, University of California, Los Angeles, CA 90095, USA}
\affiliation{Mani L. Bhaumik Institute for Theoretical Physics, University of California, Los Angeles, CA 90095, USA}
\affiliation{Center for Frontiers in Nuclear Science, Stony Brook University, Stony Brook, NY 11794, USA}

\author{Piotr Korcyl}
\affiliation{Institute of Theoretical Physics, Jagiellonian University, ul. Lojasiewicza 11, 30-348 Krakow, Poland}

\author{Farid Salazar}
\affiliation{Department of Physics, Temple University, Philadelphia, Pennsylvania 19122, USA}
\affiliation{RIKEN-BNL Research Center, Brookhaven National Laboratory, Upton, New York 11973, USA}
\affiliation{Physics Department, Brookhaven National Laboratory, Upton, New York 11973, USA}

\author{Bj\"orn~Schenke}
\affiliation{Physics Department, Brookhaven National Laboratory, Upton, New York 11973, USA}

\author{Tomasz Stebel}
\affiliation{Institute of Theoretical Physics, Jagiellonian University, ul. Lojasiewicza 11, 30-348 Krakow, Poland}

\author{Raju Venugopalan}
\affiliation{Physics Department, Brookhaven National Laboratory, Upton, New York 11973, USA}
\affiliation{CFNS, Department of Physics and Astronomy, Stony Brook University, Stony Brook, NY 11794, USA}
\affiliation{Higgs Center for Theoretical Physics, The University of Edinburgh, Edinburgh, EH9 3FD, Scotland, UK}

\author{Wenbin Zhao}
\affiliation{Institute of Particle Physics and Key Laboratory of Quark and Lepton Physics (MOE), Central China Normal University, Wuhan, 430079, Hubei, China}

\begin{abstract}

We present a detailed numerical investigation of semi-inclusive forward di-hadron production in proton–nucleus collisions employing the Color Glass Condensate effective theory. We focus on the regime where di-hadrons are produced nearly back-to-back in the transverse plane, thereby justifying a transverse-momentum-dependent factorization approach in terms of small‑$x$ gluon distributions. Our computation integrates several key elements: i) non-linear rapidity evolution via the Balitsky–Kovchegov equation with running coupling, ii) both perturbative and non-perturbative Sudakov resummation, and iii) a phenomenologically constrained model for the initial conditions for small-$x$ gluon distributions. We compare this phenomenological framework to experimental data from the STAR Collaboration on azimuthal correlations in forward di‑pion production in both proton–proton and proton–gold collisions. We analyze the systematic theoretical uncertainties associated with the saturation scales of nuclei at the initial scale for rapidity evolution and with those associated with the hadronization process. Finally, we make predictions for the kinematics anticipated to be covered by the ALICE Forward Calorimeter (FoCal) upgrade at the Large Hadron Collider.

\end{abstract}

\maketitle

\section{Introduction}
\label{sec:intro}

The discovery and characterization of gluon saturation is 
an outstanding challenge at the intersection of high-energy QCD, hadron structure, and heavy-ion physics. Gluon saturation occurs when the gluon phase-space density (or equivalently, its field strength) in a hadron or nucleus becomes sufficiently large such that recombination and screening effects compete with parton splitting contributions, with the net effect of taming the growth of gluon distributions~\cite{Gribov:1983ivg,Mueller:1985wy}. This phenomenon is expected for partons carrying small longitudinal momentum fraction $x$ and transverse momenta below an energy- and nuclear size $A$–dependent momentum scale~$Q_s(x,A)$, known as the saturation scale.

The Color Glass Condensate (CGC) effective theory provides a systematic framework to describe this regime of QCD~\cite{McLerran:1993ni,McLerran:1993ka,McLerran:1994vd,Ayala:1995kg,Ayala:1995hx}. (For reviews, see~\cite{Iancu:2003xm,Gelis:2010nm,Kovchegov:2012mbw,Albacete:2014fwa,Morreale:2021pnn}.) The highly occupied gluon system is treated as a semiclassical non-Abelian gauge background field and the scattering amplitudes of fast moving color charges propagating through this background are expressed in terms of light-like Wilson lines, which resum multiple eikonal interactions with the background field. As a result, particle-production cross sections can be expressed as convolutions of correlators of light-like Wilson lines with perturbatively calculable factors.

At high energies, leading logarithmic ($\alpha_s \ln(x_0/x)\sim O(1)$) quantum corrections become large. These are resummed into the renormalization group evolution with $x$ (starting from an initial $x_0$)  of multi-point Wilson-line correlators that satisfy the Balitsky-JIMWLK hierarchy~\cite{JalilianMarian:1996xn,JalilianMarian:1997dw,Kovner:2000pt,Iancu:2000hn,Iancu:2001ad,Ferreiro:2001qy,Weigert:2005us,Balitsky:1995ub,Kovchegov:1999yj}. In an $N_c\gg 1$, $A\gg 1$ mean-field approximation, the first equation of the hierarchy governing the evolution of the dipole correlator is the Balitsky–Kovchegov (BK) equation. The CGC EFT has been  applied to study a wide range of observables at the Hadron-Electron Ring Accelerator (HERA), the Relativistic Heavy Ion Collider (RHIC), the Large Hadron Collider (LHC), and the future Electron-Ion Collider (EIC) (see~\cite{Morreale:2021pnn}).

Measurements of di-hadron correlations, in particular the azimuthal angular distribution of hadron pairs produced in the forward region, have long been recognized as a sensitive probe of saturation dynamics in  proton–proton and proton-nucleus collisions~\cite{Kharzeev:2004bw,Marquet:2007vb,Albacete:2010pg}. The forward kinematics selects large-$x$ partons from the proton and small-$x$ gluons from the nucleus. Furthermore, if one focuses on configurations where the di-hadrons are produced nearly back-to-back (the away-side peak in the measured relative rapidity $\Delta \eta$ and relative azimuthal angle $\Delta\Phi$ plane), the transverse momentum imbalance is a sensitive probe of the nuclear saturation scale.

To characterize the correlations, it is convenient to introduce the coincidence probability: the probability of producing an associated hadron~$h_2$, conditioned on the presence of a trigger hadron~$h_1$ in another bin:
\begin{align}
CP(\Delta\Phi) = N_{\rm pair}(\Delta\Phi)/N_{\rm trig} \,.
\label{eq:coincidence-prob}
\end{align}
Here,
\begin{align}
& N_{\rm pair}(\Delta\Phi) = \int \der \eta_{h_1} \int \der \eta_{h_2} \int \der \phtone^2 \int \der \phttwo^2 \nonumber \\
& \times \frac{\der \sigma^{pA \to h_1 h_2 X}}{\der (\Delta\Phi) \der \eta_{h_1} \der \eta_{h_2} \der \phtone^2 \der \phttwo^2}
\end{align}
counts the number of hadron pairs in the specified kinematic bins and with relative azimuthal angle~$\Delta\Phi$, while
\begin{align}
& N_{\rm trig} = \int \der \eta_{h_1} \int \der \phtone^2 \, \frac{\der \sigma^{pA \to h_1 X}}{\der \eta_{h_1} \der \phtone^2}
\end{align}
counts the number of trigger particles in a rapidity and transverse momentum bin.

In the away-side region ($\Delta\Phi \sim \pi$), saturation is expected to yield two key effects: broadening and suppression. Broadening arises from multiple scatterings in the classical gluon field, while suppression is induced by non-linear quantum evolution~\cite{Kharzeev:2003wz,Albacete:2003iq}. These effects are more pronounced with increasing saturation scale, which grows with collision energy, increasingly forward rapidities (smaller~$x$), larger nuclear targets, and/or more central collisions. Di-$\pi^0$ correlations were studied within the CGC formalism (including at leading order only the quark-initiated channel), for a Gaussian distribution of color sources \cite{Lappi:2012nh}.
The differential cross-section is expressed as a convolution of the light-cone wavefunction for the splitting of a quark into quark+gluon pair and correlators of multi-point correlators of Wilson lines. The results in \cite{Lappi:2012nh} predict both broadening and suppression of $CP(\Delta\Phi)$. 

In \cite{Dominguez:2011wm,Dominguez:2010xd}, it was shown that in the back-to-back kinematics, the dominant contribution to the differential cross section factorizes into a hard scattering coefficient and a nonperturbative small-$x$ transverse-momentum–dependent parton distribution functions (TMDs). In the CGC, these TMDs are expressed through combinations of Wilson lines and their derivatives, with their rapidity evolution governed by the BK–JIMWLK equations. (For numerical studies, see \cite{Marquet:2016cgx,Petreska:2018cbf,Cali:2021tsh}.) The gauge link structure of each TMDs reflects the color flow of the underlying partonic subprocess \cite{Bomhof:2006dp}. In the CGC, the gauge link structure of the TMD is inherited from the color structure of correlators of Wilson lines. At large $N_c$ all unpolarized gluon TMDs at small-$x$ can be computed in terms of two universal objects: the dipole and  Weiszäcker-Williams correlators \cite{Dominguez:2012ad}.

Several phenomenological studies using this small-$x$ TMD framework have been carried out for proton–nucleus collisions~\cite{Jalilian-Marian:2012wwi,Stasto:2011ru,Kotko:2017oxg,Albacete:2018ruq,vanHameren:2019ysa,vanHameren:2020rqt,Benic:2022ixp,Al-Mashad:2022zbq,vanHameren:2023oiq,Ganguli:2023joy} and deep inelastic scattering~\cite{Zheng:2014vka,vanHameren:2021sqc}. In particular, Ref.~\cite{Albacete:2018ruq} studied $CP(\Delta\Phi)$ including both quark- and gluon-initiated channels with TMDs computed in the Gaussian approximation, again finding broadening and suppression. While these results agree with the observed suppression between $pp$ and $d$Au at RHIC, they do not reproduce the width of the correlation: STAR~\cite{Braidot:2010zh,STAR:2021fgw} and PHENIX~\cite{PHENIX:2011puq} data show similar widths in both systems and larger widths than predicted by leading order (LO) CGC evolution. Similarly, LHC dijet measurements~\cite{ATLAS:2019jgo} confirm strong suppression of the away-side peak but no clear nuclear broadening. Such observations, recently motivated, alternative approaches to explain the two-particle correlation data, see e.g. a nuclear PDF-based approach in \cite{Perepelitsa:2025qpz}.

A plausible explanation for the lack of broadening is that, in back-to-back kinematics, parton showering and intrinsic transverse momentum broadening from fragmentation dominate the width of the correlation. A recent study~\cite{Cassar:2025vdp} found that including parton showers and fragmentation smearing renders the width of $CP(\Delta\Phi)$ insensitive to variations in the intrinsic $k_T$ of the small-$x$ gluon, suggesting that saturation plays only a subleading role in controlling the width of the away-side peak.

The small-$x$ TMD formalism provides a natural way to incorporate these competing mechanisms. In back-to-back kinematics, higher-order corrections are enhanced not only by small-$x$ logarithms but also by large Sudakov double and single logarithms. At moderate~$x$, the Collins–Soper–Sterman (CSS) equations resum these logarithms \cite{Collins:1981uk,Collins:1981uw,Collins:1984kg,Collins:2011zzd,Gao:2023ulg,Sun:2015doa,Kang:2020xez}. Joint resummation of small-$x$ and Sudakov logarithms was proposed in~\cite{Mueller:2012uf,Mueller:2013wwa,Zhou:2018lfq} and has since been extensively studied~\cite{Xiao:2017yya,Hatta:2020bgy,Hatta:2021jcd,Zhou:2018lfq,Hentschinski:2021lsh,Taels:2022tza,Caucal:2022ulg,Caucal:2023nci,Caucal:2023fsf,Caucal:2024bae,Caucal:2024vbv,Caucal:2024nsb,Caucal:2025mth,Duan:2024nlr,Duan:2024qev,Mukherjee:2023snp}. Perturbative Sudakov logarithms are largely captured by parton showers, while the nonperturbative Sudakov factor can be interpreted as intrinsic transverse momentum from fragmentation. Including these effects is therefore essential for correctly describing the width of the away-side correlation and addressing the puzzle of missing broadening. Ref.~\cite{Stasto:2018rci} incorporated Sudakov resummation into the small-$x$ TMD framework using a simple saturation model with parametrized $x$ and $A$ dependence, demonstrating little variation in widths of the self-normalized differential cross-section when comparing $pp$ and $pA$ collisions; hence, demonstrating the lack of broadening. On the other hand, because the correlation was self-normalized, information about suppression was lost, limiting insight into saturation dynamics.

 In this paper, we present a computation of di-hadron correlations in the CGC within the improved TMD framework that includes Sudakov resummation, small-$x$ evolution, and phenomenologically constrained initial conditions for the gluon distributions in $pp$ and $pA$ collisions. We focus on the away-side peak, where the near–back-to-back kinematics justifies a TMD factorization in terms of small-$x$ gluon distributions. Within a Gaussian approximation, we relate all small-$x$ TMDs to the Wilson-line dipole correlator and evolve it using the running-coupling BK equation \cite{Balitsky:2006wa}. Both perturbative and nonperturbative Sudakov logarithms are included, and we employ off-shell hard factors from the improved TMD formalism~\cite{Kotko:2015ura,Altinoluk:2019fui}. Our initial proton and nuclear dipole distributions, respectively, are drawn from fits to HERA reduced cross-section data for the proton and from a minimum-bias nuclear dipole model incorporating the transverse nuclear density~\cite{Deganutti:2023qct}. We compare our results to STAR data on forward di-hadron azimuthal correlations in $pp$ and pAu collisions~\cite{STAR:2021fgw}. We also provide a detailed study of theoretical uncertainties associated with the nuclear initial saturation scale and hadronization. Finally, we present predictions for the kinematics expected to be accessible with the ALICE FoCal upgrade at the LHC~\cite{ALICE-PUBLIC-2019-005}.

The paper is organized as follows. Section~II reviews the computation of forward di-hadron production in $pA$ collisions within the small-$x$ TMD formalism. Section~III details the nonperturbative inputs used in our calculation, including our choice for the initial conditions for the dipole correlator, parton distribution functions, fragmentation functions, and the nonperturbative Sudakov factor. Section~IV presents numerical results compared with experimental data from STAR, along with predictions for FoCal. We conclude in Section~V with a summary and outlook for future measurements at RHIC, the LHC, and beyond.

\section{Theoretical formalism}

\label{sec:theory-sec}

In the eikonal approximation, the multiple scattering of quarks and gluons with the small-$x$ gluon background field is encoded in their color rotation via light-like Wilson lines in the fundamental and adjoint representations, respectively. They are given by
\begin{align}
    V(\xt) &= \mathcal{P}\exp\left[ ig \int_{-\infty}^{\infty} \der x^+ A_a^-(x^+,\xt) t^a \right] \,, \label{eq:WilsonF}\\
    U(\xt) &= \mathcal{P}\exp\left[ ig \int_{-\infty}^{\infty} \der x^+ A_a^-(x^+,\xt) T^a \right]  \,,
    \label{eq:WilsonA}
\end{align}
where $\mathcal{P}$ denotes path ordering on the exponential of the color matrix. $t^a$ and $T^a$ are the generators of SU(3) in the fundamental and adjoint representations, respectively. The background field, in $A^+ =0$ gauge, obeys
\begin{align}
    \nabla_\perp^2 A_a^-(x^+,\xt) = - \rho^a(x^+,\xt) \, \quad A_\perp^i = 0\,, \quad A^+ =0 \,,
\end{align}
where $\rho^a(x^+,\xt)$ is the color charge density of classical sources representing the large-$x$ partons. 

Differential cross-sections are then expressed in terms of multi-point operators $\Ocal$ of Wilson lines, averaged over 
configurations of color sources $\rho^a$. The two simplest such operators are the dipoles in fundamental and adjoint representations \cite{Jalilian-Marian:2004vhw,Baier:2005dv}
\begin{align}
    S_F^{(2)}(\bt,\bt') &= \frac{1}{N_c} \Tr\left[ V(\bt) V^\dagger(\bt') \right]\,, \\
    S_A^{(2)}(\bt,\bt') &= \frac{1}{N_c^2-1} \Tr\left[ U(\bt) U^\dagger(\bt') \right]\,.
\end{align} 
The average over $\rho^a$ is denoted as
\begin{align}
    \langle \Ocal \rangle_x = \int [\mathcal{D} \rho] W_x[\rho] \Ocal[\rho]
\end{align}
where $W_x[\rho]$ is a gauge invariant stochastic weight functional describing the distribution of sources, whose dependence on $x$ is governed by the JIMWLK equation \cite{JalilianMarian:1996xn,JalilianMarian:1997dw,Kovner:2000pt,Iancu:2000hn,Iancu:2001ad,Ferreiro:2001qy,Weigert:2005us}.

Let us now consider the single inclusive production of a hadron in proton-nucleus collision. A hadron in the forward region (proton-going) is produced from a large-$x$ parton in the proton that scatters off the low-$x$ gluon background field of the nucleus. In these asymmetric kinematics, large-$x$ partons in the proton can be treated as collinear, and thus described via parton distribution functions (PDFs). The scattering with the low-$x$ gluon field is encoded in light-like Wilson lines (see Eqs.\, \eqref{eq:WilsonF} and \eqref{eq:WilsonA}). This setup~\cite{Dumitru:2005gt} can be obtained systematically within the so-called dilute-dense approximation of the CGC EFT~\cite{Gelis:2003vh,Blaizot:2004wu,Blaizot:2004wv}. The differential cross-section at leading order in the dilute-dense framework reads~\cite{Dumitru:2005gt}
\begin{widetext}
    \begin{align}
    \frac{\der \sigma^{pA \to h_1 X} }{\der^2 \phtone \der \eta_{h_1} } = \int_{z_{h_{1,\rm min}}}^1 \frac{\der z_{h_1}}{z^2_{h_1}} \left[ \sum_{q=u,d} x_p f_q(x_p,\mu^2) N_{\rm F}(x_g,\Kt) D_{h_1/q}(z_{h_1},\mu^2) + x_p f_g(x_p,\mu^2) N_{\rm A}(x_g,\Kt) D_{h_1/g}(z_{h_1},\mu^2) \right],
        \label{eq:single-inclusive}
\end{align}
\end{widetext}
with $\Kt=\phtone/z_{h_1}$, with magnitude $K_\perp = |\Kt|$ \footnote{Throughout the text we will denote the magnitude of transverse vectors as $v_\perp = |\boldsymbol{v}|$.}. The longitudinal momentum fractions are given by $x_{p} = K_\perp e^{\eta_1} /\sqrt{s}$, and $x_{g} = K_\perp e^{- \eta_1} /\sqrt{s}$, and we assume the partonic rapidity $\eta_1$ to be equal to the hadronic rapidity $\eta_{h_1}$. The bound $z_{h_{1,\rm min}}$ ensures the energy constraint $x_{p}, x_{g}<1$. The factorization scale is chosen $\mu^2 = \Kt^2$. Here, $f_q$ and $f_g$ are the quark and gluon PDFs, and $D_{h/q}$ and $D_{h/g}$ are the fragmentation functions into a hadron from a quark or gluon, respectively. We defined the Fourier transform of the dipole amplitude in fundamental and adjoint representations, respectively
\begin{align}
    N_{\rm F}(x_g,\Kt) &= \int \der^2 \xt \ \der^2 \xt' e^{-i \Kt \cdot (\xt-\xt')} \langle S_F^{(2)}(\xt,\xt') \rangle_{x_g} \,,\\
    N_{\rm A}(x_g,\Kt) &= \int \der^2 \xt \ \der^2 \xt' e^{-i \Kt \cdot (\xt-\xt')} \langle S_A^{(2)}(\xt,\xt') \rangle_{x_g} \,. 
\end{align}
Similarly, dihadrons produced in the forward region arise from a large-$x$ parton in the proton which radiates a second parton before or after the scattering off the low-$x$ background field. If the hadron imbalance is small, i.e., back-to-back kinematics, the scattering effectively occurs off a small-$x$ gluon, whose distribution is described by an appropriate TMD \cite{Dominguez:2011wm}. The hard scattering is described by an on-shell hard coefficient, and in the improved TMD formalism by off-shell hard factors \cite{Kotko:2015ura,Marquet:2016cgx}. We will employ the latter as it will allow us to cover a larger angular window around $\Delta\Phi \sim \pi$ without resulting in much more numerical complexity.

The differential cross-section for forward di-hadron production in proton–nucleus collisions is as a result the convolution of the leading-order parton cross-section, the Sudakov factor, the collinear parton distribution functions (PDFs) and the fragmentation functions (FFs). Including quark and gluon-initiated channels, it reads
\begin{widetext}
\begin{align}
    & \frac{\der \sigma^{pA \to h_1 h_2 X}}{\der \eta_{h_1} \der \eta_{h_2} \der^2 \phtone \der^2 \phttwo} = \int_{z_{h_{1,\rm min}}}^1 \frac{\der z_{h_1}}{z_{h_1}^2} \int_{z_{h_{2,\rm min}}}^1 \frac{\der z_{h_2}}{z_{h_2}^2} \int \frac{\der^2 \kt'}{(2\pi)^2} \Bigg\{ W_{gg \to gg}(z_{h_1},z_{h_2},x_p,\kt-\kt',Q^2) \nonumber \\
    & \times \left[ (\Fcal_{gg}^{(1)}(x_g,\kt') + \Fcal_{gg}^{(3)}(x_g,\kt') ) \Hcal^{(1)}_{gg \to gg}(z,\ptone,\pttwo) +   (\Fcal_{gg}^{(2)}(x_g,\kt') + \Fcal_{gg}^{(3)}(x_g,\kt') )  \Hcal^{(2)}_{gg \to gg}(z,\ptone,\pttwo) \right] \nonumber \\
    &  + W_{qg \to qg}(z_{h_1},z_{h_2},x_p,\kt-\kt',Q^2) \left[ \Fcal_{qg}^{(1)}(x_g,\kt') \Hcal^{(1)}_{qg \to qg}(z,\ptone,\pttwo) +   \Fcal_{qg}^{(2)}(x_g,\kt') \Hcal^{(2)}_{qg \to qg}(z,\ptone,\pttwo)  \right]  \nonumber \\
    & + W_{qg \to qg}(z_{h_2},z_{h_1},x_p,\kt-\kt',Q^2) \left[  \Fcal_{gg}^{(1)}(x_g,\kt') \Hcal^{(1)}_{gg \to q\bar{q}}(z,\ptone,\pttwo) +  \Fcal_{gg}^{(2)}(x_g,\kt')  \Hcal^{(2)}_{gg \to q\bar{q}}(z,\ptone,\pttwo) \right]  \nonumber \\
    & + 2  W_{gg \to q\bar{q}}(z_{h_1},z_{h_2},x_p,\kt-\kt',Q^2) \left[\Fcal_{gg}^{(1)}(x_g,\kt') \Hcal^{(1)}_{gg \to q\bar{q}}(z,\ptone,\pttwo) +  \Fcal_{gg}^{(2)}(x_g,\kt')  \Hcal^{(2)}_{gg \to q\bar{q}}(z,\ptone,\pttwo) \right] \Bigg \}\,,
    \label{eq:double-inclusive}
\end{align}
\end{widetext}
where $\phtk$ and $\eta_{h_k}$ denote the transverse momenta and rapidities of the final state hadrons, respectively. The transverse momenta of the partons that fragment into the final state hadrons are $\ptk = \phtk /z_{h_k}$, and parton and hadron rapitidies are assumed to be equal $\eta_{h_i} = \eta_i$. Furthermore, $\kt = \ptone + \pttwo$ is the partonic momentum imbalance, and $z = p_{1\perp} e^{\eta_1}/(p_{1\perp} e^{\eta_1} + p_{2\perp} e^{\eta_2})$ is the longitudinal momentum of one the final state partons reltive to the incoming parton from the proton. The longitudinal momentum fractions probed are $x_{p} = (p_{1\perp} e^{\eta_1} + p_{2\perp} e^{ \eta_2})/\sqrt{s}$ and $x_{g} = (p_{1\perp} e^{-\eta_1} + p_{2\perp} e^{-\eta_2})/\sqrt{s}$, and the bounds $z_{h_{i,\rm min}}$ ensure the energy constraint $x_{p}, x_g <1$. The functions $\Fcal^{(n)}_{\alpha}(x_g,\kt)$ denote the gluon TMDs; while the functions $W_{ij \to kl}(z_{h_1},z_{h_2},x_p,\kt-\kt',Q^2)$ account for the Sudakov factor, the PDFs and the FFs. The off-shell gluon hard factors are denoted by $\Hcal(z,\ptone,\pttwo)$, and their complete expressions are given in Appendix \ref{app:ITMD-hard-factor}. The integration over $\kt'$ convolves the gluon TMDs with the Sudakov factors. This setup is equivalent to the conventional implementation of the Sudakov factor in position space. Lastly, we choose the factorization scale $Q^2 = x_p x_g s$.

The small‑$x$ gluon TMDs, $\Fcal^{(n)}_{\alpha}(x_g,\kt)$, are constructed from various combinations of Wilson lines and their derivatives \cite{Dominguez:2011wm}. These combinations encapsulate the distinct gauge link structures, which are dictated by the color flow of the underlying channels. In the CGC EFT, they are given by
\begin{align}
    \Fcal^{(n)}_{\alpha}(x_g,\kt) = \int \der^2 \bt\  \der^2 \bt' e^{-i \kt \cdot(\bt-\bt')} \left \langle \widetilde{\Fcal}^{(n)}_{\alpha}(\bt,\bt')\right \rangle_{x_g}\,, \label{eq:CGC-TMD-def}
\end{align}
with $\alpha \in \{qg, gg\}$, and
\begin{align}
    \widetilde{\Fcal}_{qg}^{(1)} &= \frac{2}{\alpha_s}  \Tr\left[ \left(\partial_\perp^i V(\bt) \right) \left( \partial_\perp^i V^\dagger(\bt') \right) \right] \nonumber, \\
    \widetilde{\Fcal}_{qg}^{(2)} &= \frac{2}{\alpha_s}   \Tr\left[ A^i_\perp(\bt) A^i_\perp(\bt') \right] S_F^{(2)}(\bt,\bt'), \nonumber \\
    \widetilde{\Fcal}_{gg}^{(1)}&= \frac{2}{\alpha_s}  \Tr\left[\left( \partial_\perp^i V(\bt) \right) \left( \partial_\perp^i V^\dagger(\bt') \right) \right] S_F^{(2)}(\bt,\bt'), \nonumber \\
    \widetilde{\Fcal}_{gg}^{(2)}&= \frac{-2}{\alpha_s} \frac{\Tr\left[\left( \partial_\perp^i V^\dagger(\bt) \right) V(\bt')  \right]}{N_c}  \Tr\left[V(\bt) \left(\partial^i_\perp V^\dagger(\bt') \right)\right],  \nonumber\\
    \widetilde{\Fcal}_{gg}^{(3)}&= \frac{2}{\alpha_s}   \Tr\left[ A^i_\perp(\bt) A^i_\perp(\bt') \right] S_F^{(2)}(\bt,\bt') S_F^{(2)}(\bt,\bt'), \label{eq:tmds}
\end{align}
where we introduced the transverse gauge field
\begin{align}
    A^k_\perp(\bt) = \frac{i}{g} V(\bt)\partial_\perp^k V^\dagger(\bt) \,,
\end{align}
which corresponds to the transverse gauge field at $x^+=+\infty$ in the gauge $A^-=0$, and $\partial_\perp^k$ is a shorthand for the partial derivative with respect to $\bt^k$.

The $x$-dependence of the gluon TMDs follows the JIMWLK evolution equation. They have been studied numerically in \cite{Marquet:2016cgx,Petreska:2018cbf,Cali:2021tsh}. In this work, the TMDs are calculated as in~\cite{Cougoulic:2024jnd}, using a Gaussian and mean-field approximations to express the TMDs in terms of the dipole amplitude, $\langle S^{(2)}(\bt,\bt') \rangle_x$, see Appendix \ref{app:TMDs-Gaussian}. The $x$-dependence of the dipole then follows the running coupling BK evolution equation with initial conditions specified in the next section.

We now detail the functions $W_{ij \to kl}(z_1,z_2,x,\lt,Q^2)$ that absorb the Sudakov factor, the PDFs and FFs. These are given by
\begin{align}
    & W_{qg \to qg}(z_1,z_2,x, \lt,Q^2) = \sum_{q=u,d} \int \der^2 \bt\ e^{-i \lt \cdot \bt} \nonumber \\
    &  x f_q(x,\mu_b^2) D_{h_1 /g}(z_1,\mu_b^2) D_{h_2 /q}(z_2,\mu_b^2) e^{-S_{qg\to q g}(b_\perp,Q^2)} \,,
\end{align}
\begin{align}
    & W_{gg \to q\bar{q}}(z_1,z_2,x,\lt,Q^2) = \sum_{q=u,d,s} \int \der^2 \bt\ e^{-i \lt \cdot \bt} \nonumber \\
    &  x f_g(x,\mu_b^2) D_{h_1/q}(z_1,\mu_b^2) D_{h_2 /q}(z_2,\mu_b^2)  e^{-S_{gg\to q \bar{q}}(b_\perp,Q^2)} \,,
\end{align}
\begin{align}
    & W_{gg \to gg}(z_1,z_2,x,\lt,Q^2) = \int \der^2 \bt\ e^{-i \lt \cdot \bt} \nonumber \\
    &  x f_g(x,\mu_b^2) D_{h_1/g}(z_1,\mu_b^2) D_{h_2 /g}(z_2,\mu_b^2)  e^{-S_{gg\to gg}(b_\perp,Q^2)} \,,
\end{align}
where the factorization scale is chosen as $\mu_b = 2 e^{-\gamma_E}/b^{*}_\perp$ with the prescription $b_\perp^* = b_\perp/\sqrt{1 + b_\perp^2/b_{\rm max}^2}$~\cite{Collins:1984kg} to control the non-perturbative region $b_\perp > b_{\rm max}$, with $b_{\rm max} = 1.5\ {\rm GeV^{-1}}$.

The Sudakov factor is given by the sum of perturbative and non-perturbative contributions,
\begin{widetext}
    \begin{align}
    S_{q g \to qg}(b_\perp,Q^2) & = \int_{\mu_b^2}^{Q^2} \frac{\der \mu^2}{\mu^2} \frac{\alpha_s(\mu)}{2\pi} \left[ \left(2 C_A + 2 C_F \right) \ln\left( \frac{Q^2}{\mu^2} \right) - \left( 3 C_F + 2 \beta_0 \right) \right] + C_{NP}\times S_{qg\to qg}^{\rm{NP}}(b_\perp,Q^2)\label{eq:sqgqg} \,, \\
    S_{g g \to gg}(b_\perp,Q^2) & = \int_{\mu_b^2}^{Q^2} \frac{\der \mu^2}{\mu^2} \frac{\alpha_s(\mu)}{2\pi} \left[ 4 C_A \ln\left( \frac{Q^2}{\mu^2} \right) -  6 \beta_0  \right] + C_{NP}\times S_{gg\to gg}^{\rm{NP}}(b_\perp,Q^2) \label{eq:sgggg} \,,\\
    S_{g g \to q\bar{q}}(b_\perp,Q^2) &= S_{q g \to qg}(b_\perp,Q^2) \,,
\end{align}
\end{widetext}
where color Casimirs are $C_A = N_c$, $C_F = (N_c^2-1)/(2N_c)$, and $\beta_0 = (11 C_A - 2N_f)/12$ with $N_f=3$. We note that, in the Sudakov factors, the single logarithm proportional to $\beta_0$ associated with small-$x$ gluons is not included. Indeed, as shown
in Ref.~\cite{Mueller:2012uf,Mueller:2013wwa,Caucal:2023nci}, it is equivalent either (i) to include this term by evaluating one factor of $\alpha_s$ in the hard factor at the hard scale
$Q$, while the factor $1/\alpha_s$ in the gluon TMDs is evaluated at the scale $\mu_b$, or (ii) to omit this single logarithm and directly cancel one
(assumed fixed) factor of $\alpha_s$ in the hard factor against a factor $1/\alpha_s$ in the TMD. The latter option is adopted here. Following \cite{Stasto:2018rci}, we include a multiplicative factor $C_{NP}$ to the non-perturbative factor, reflecting the non-universality between extractions from SIDIS and Drell-Yan, as compared to dihadron production in hadronic collisions. The parametrization for the non-perturbative Sudakov will be discussed in the next section.

The coincidence probability can now be computed from Eqs.\,\eqref{eq:double-inclusive} and \eqref{eq:single-inclusive}.

\section{Non-perturbative input}

\label{sec:np-input}

In our numerical study, the PDFs are taken from the CT18NNLO fits \cite{Hou:2019qau}, and the default fragmentation functions are taken from the NLO JAM19 analysis \cite{Sato:2019yez}. In addition, we consider the NP23 \cite{Gao:2024nkz} and DSS07 \cite{deFlorian:2007aj} fragmentation functions to test the sensitivity of our results to final state fragmentation effects. 

We employ a phenomenologically constrained dipole amplitude that has been fitted to the deep inelastic scattering reduced cross-section at HERA \cite{Albacete:2010sy} (see also \cite{Lappi:2013zma}). In such fit the dipole is impact-parameter independent, i.e., it only depends on the difference of coordinates $\langle S_F^{(2)}(\bt-\bt') \rangle_x$. 
The dipole amplitude initial condition is set at $x_0 = 0.01$ using the modified McLerran–Venugopalan (MV) model with an anomalous dimension, referred to as the MV$^\gamma$ model. The initial condition for the evolution of the fundamental dipole amplitude is given by
\begin{equation}
\calN(x_0,r_\perp) =1 - \exp\left\{- \frac{\left(r_\perp^2 Q_{s0}^2\right)^{\gamma}}{4} \log\left(\frac{1}{r_\perp \Lambda_{\rm IR}}+ e\right)\right\}\,,
\end{equation}
where $\calN(x_0,r_\perp)  = 1- \langle S^{(2)}_F(r_\perp)\rangle_{x_0}$ is the initial fundamental dipole amplitude, $r_\perp = |\bt-\bt'|$ is the transverse size of the dipole, $Q_{s0}$ is the initial saturation momentum at $x_0 = 0.01$, $\gamma$ is the anomalous dimension and $\Lambda_{\rm IR}$ is the IR cutoff suppressing Coulomb tails. We use the parameter set $\gamma = 1.119$, $Q^2_{s0,p} = 0.168 \, {\rm GeV}^2$, $\Lambda_{\rm IR} = 0.241 \, {\rm GeV}$. The running coupling in the rcBK is 
\begin{align}
    \alpha_{s}(r^2) = \frac{\pi}{\left(\beta_0 \ln \left( \frac{4C^2}{r^2 \Lambda_{\rm IR}^2} \right)\right)} \,,
\end{align} 
with $C = 1.715$.
Such set-up provides the best fit to the inclusive structure functions and reduced cross sections measured in electron-proton scattering \cite{Albacete:2010sy}.  The rcBK equation evolves the dipole distribution to values $x < 0.01$. However, the MV$^\gamma$ initial condition can lead to spurious negative TMDs after Fourier transform to momentum space for values of $x$ close to $x_0$. This is an artifact of the initial condition which is then removed after a unit of rapidity in evolution. To address this, we employ the TMDs obtained from the solution rcBK evolution equation for $x < x_c$ (with $x_c = x_0 e^{-1}= 0.0037$) and perform a matching for $x > x_c$ using the prescription \cite{Gelis:2006tb}
\begin{equation}
\Fcal^{(i)}_{\alpha}(x,\kt) = \left(\frac{1-x}{1-x_c} \right)^4 \Fcal^{(i)}_{\alpha}(x_c,\kt) \,.
\end{equation}
The dipole in the adjoint representation is obtained using $\langle S^{(2)}_A(r_\perp)\rangle_{x} = \left( \langle S^{(2)}_F(r_\perp)\rangle_{x} \right)^2$.

For nuclei, we follow the procedure outlined in Ref.~\cite{Deganutti:2023qct}, which relates the saturation scales for minimum bias events in protons and nuclei. This procedure effectively accounts for the transverse gluon density in protons and nuclei with a Gaussian and Woods-Saxon distribution, respectively. In particular, we use $Q_{s0, {\rm Au}}^2 = 2.67 \ Q^2_{s0,p}$ and $Q_{s0, {\rm Pb}}^2 = 2.75 \ Q^2_{s0,p}$ (similar scalings of the saturation scale at $x=0.01$ can be inferred from \cite{Kowalski:2007rw}).  Due to uncertainties in the determination of the nuclear initial scale, we vary $Q_{s0,{\rm Pb/Au}}^2$ by approximately 20\% to account for uncertainties arising from the modeling of the impact-parameter dependence of the initial condition. 

\begin{figure*}
    \centering    \includegraphics[width=1.8\columnwidth]{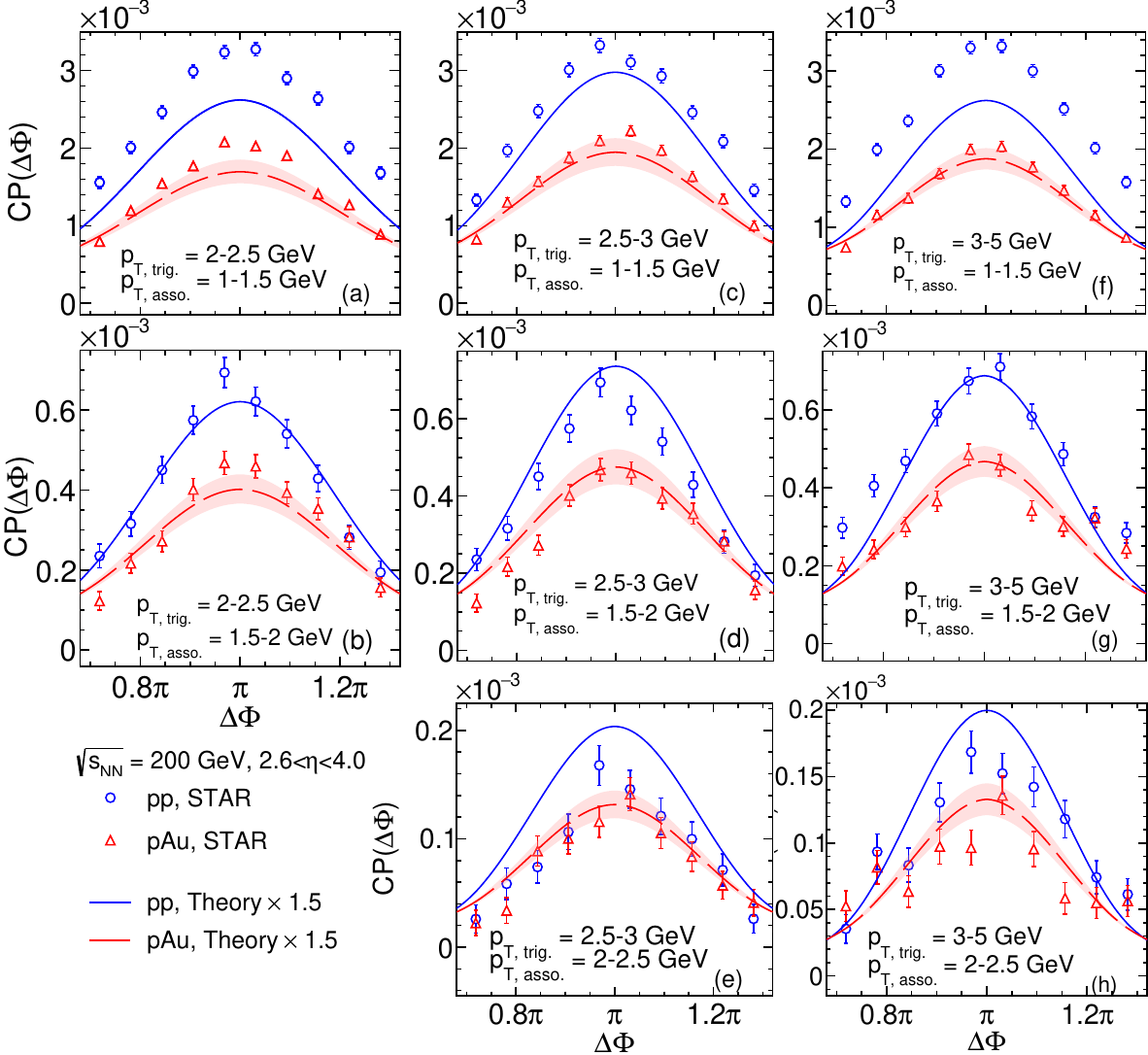}
    \caption{Coincidence probability in the away-side region of di-$\pi^0$ production in p-p and p-Au collisions at $\sqrt{s_{NN}}=200$ GeV, $2.6<\eta<4.0$ for various $p_T$ ranges. The band is obtained by changing the $Q^2_{\rm s, Au}$ by a factor of 1.2. The STAR data is from \cite{STAR:2021fgw}.}
    \label{fig:star_8_panel}
\end{figure*}

For the non-perturbative Sudakov factor, we employ the parametrization from Refs.\, \cite{Sun:2014dqm,Prokudin:2015ysa,Echevarria:2020hpy},
\begin{align}
S^{NP}_{qg\to qg} (b_\perp,Q^2) & = \left( 2 + \frac{C_A}{C_F} \right) \frac{g_1}{2} b_\perp^2 \nonumber \\ 
&+ \left(2 + \frac{C_A}{C_F} \right) \frac{g_2}{2} \ln \frac{Q}{Q_0} \ln \frac{b_\perp}{b_\perp^*}
\end{align}
\begin{align}
S^{NP}_{gg\to gg} (b_\perp,Q^2) & = \frac{3C_A}{C_F} \frac{g_1}{2} b_\perp^2 \nonumber \\ 
&+ \frac{3C_A}{C_F} \frac{g_2}{2} \ln \frac{Q}{Q_0} \ln \frac{b_\perp}{b_\perp^*}\,,
\end{align}
with $g_1 = 0.212$, $g_2 = 0.84$, and  $Q_0^2 = 2.4 \ {\rm GeV}^2$. These expressions do not include the non-perturbative part of the small-$x$ gluon TMDs which is already accounted for in the CGC small-$x$ TMDs.
Following \cite{Stasto:2018rci}, we include a multiplicative factor, $C_{NP}$,  which we set to 2.3 in Eqs.~\eqref{eq:sqgqg} and (\ref{eq:sgggg})  to fit the width for the dihadron coincidence probability $CP(\Delta\Phi)$ in p-p collisions. We refer the reader to Section \ref{sec:Results} for further details. 

\begin{figure*}
    \centering
    \includegraphics[width=2.05\columnwidth]{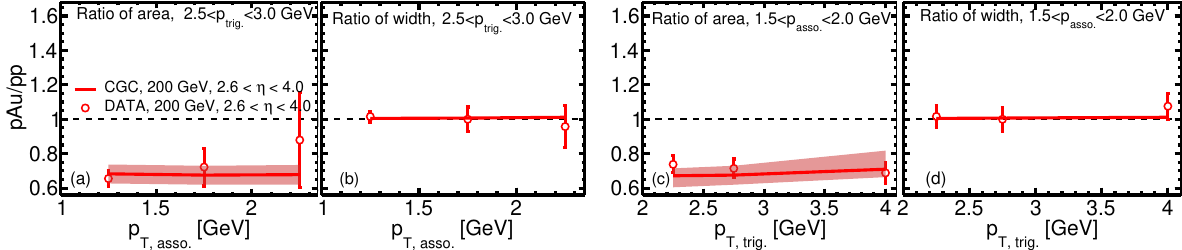}
    \caption{Relative area and width of back-to-back di-$\pi^0$ correlations at forward pseudorapidities in p-Au relative to p-p collisions for $p_{T,\rm trig}$ = 2.5 - 3 GeV/c as a function of $p_{T,\rm asso}$ (left two panels) and  for $p_{T,\rm asso}$ = 1.5 - 2 GeV/c as a function of $p_{T,\rm trig}$ (right two panels).  The band is obtained by changing the $Q^2_{s,\rm Au}$ by a factor of 1.2. The STAR data points are obtained by performing a Gaussian fit to the STAR measurements~\cite{STAR:2021fgw}. }
    \label{fig:ratio_com_star}
\end{figure*}

\begin{figure*}
    \centering
    \includegraphics[width=2.05\columnwidth]{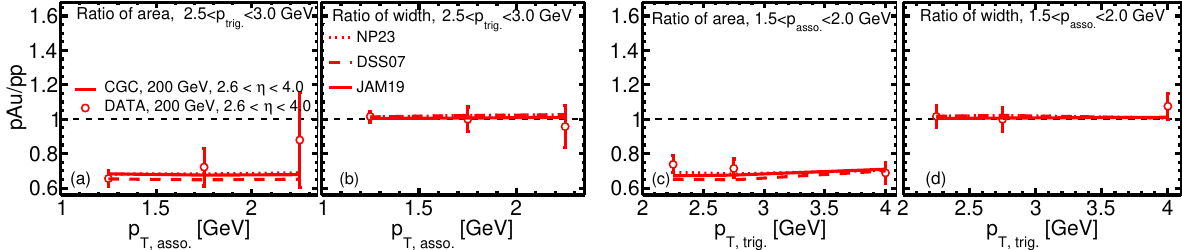}
    \caption{Relative area and width of back-to-back di-$\pi^0$ correlations at forward pseudorapidities  with three different fragmentation functions  in p-Au  relative to p-p collisions for $p_{T,\rm trig}$ = 2.5 - 3 GeV/c as a function of $p_{T,\rm asso}$ (left two panels) and  for $p_{T,\rm asso}$ = 1.5 - 2 GeV/c as a function of $p_{T,\rm trig}$ (right two panels).  The STAR data points are obtained by performing a Gaussian fit to the STAR measurements~\cite{STAR:2021fgw}.  }
    \label{fig:ratio_ff_com_star}
\end{figure*}

\begin{figure*}
    \centering
\includegraphics[width=1.8\columnwidth]{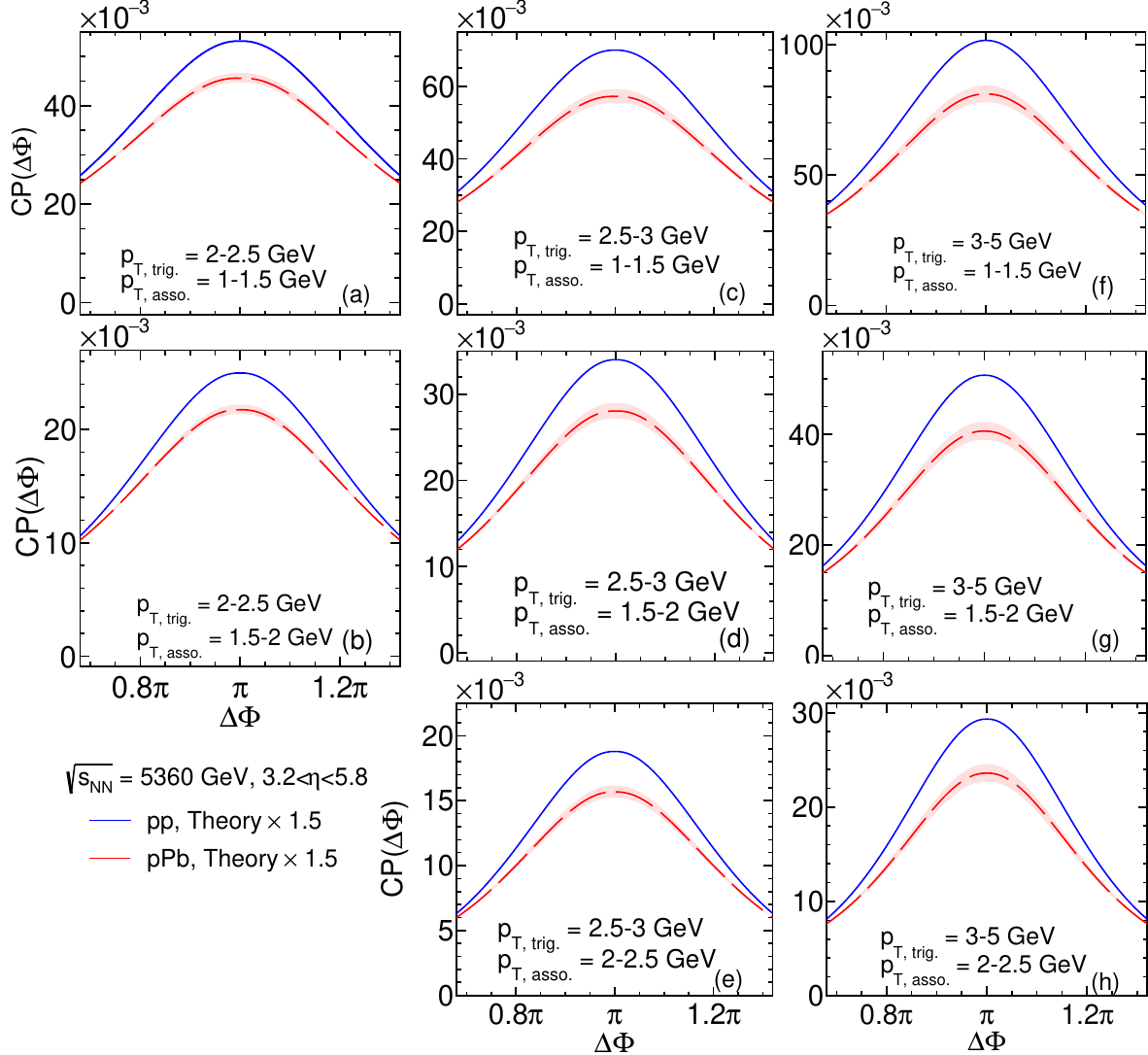}
    \caption{Coincidence probability in the away-side region of di-$\pi^0$ production in p-p and p-Pb collisions at $\sqrt{s_{NN}}=5.36$ TeV, $3.2<\eta<5.8$ for various $p_T$ ranges. The band is obtained by changing the $Q^2_{\rm s, Pb}$ by a factor of 1.2. }
    \label{fig:FoCal}
\end{figure*}

\begin{figure*}
    \centering
    \includegraphics[width=2.05\columnwidth]{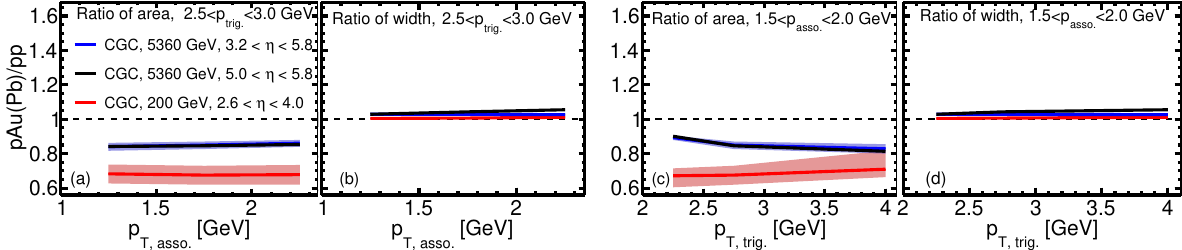}
    \caption{Relative area and width  of back-to-back di-$\pi^0$ correlations at forward pseudorapidities in p-Au  and p-Pb relative to p-p collisions for $p_{T,\rm trig}$ = 2.5 - 3 GeV/c as a function of $p_{T,\rm asso}$ (left two panels) and  for $p_{T,\rm asso}$ = 1.5 - 2 GeV/c as a function of $p_{T,\rm trig}$ (right two panels).  The band is obtained by changing the $Q^2_{s,\rm Au/Pb}$ by a factor of 1.2. }
    \label{fig:ratio_com}
\end{figure*}

\begin{figure*}
    \centering
    \includegraphics[width=2.05\columnwidth]{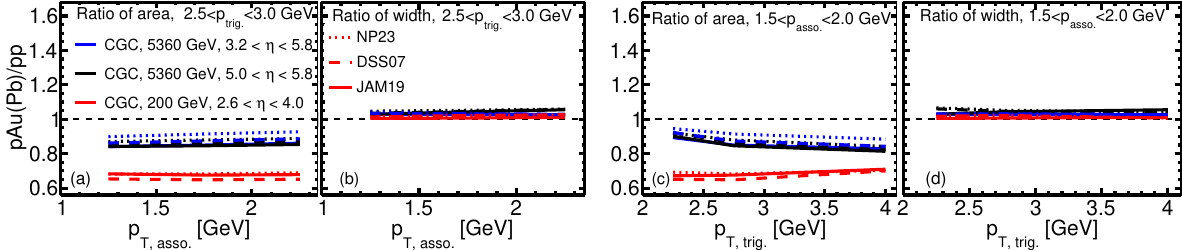}
    \caption{Relative area and width of back-to-back di-$\pi^0$ correlations at forward pseudorapidities  with three different fragmentation functions  in p-Au  and p-Pb relative to p-p collisions for $p_{T,\rm trig}$ = 2.5 - 3 GeV/c as a function of $p_{T,\rm asso}$ (left two panels) and  for $p_{T,\rm asso}$ = 1.5 - 2 GeV/c as a function of $p_{T,\rm trig}$ (right two panels). }
    \label{fig:ratio_ff_com}
\end{figure*}

\section{Results}
\label{sec:Results}

In this section, we will present the numerical results for the coincidence probability $CP(\Delta\Phi)$ for dipion production following the setup in Secs.\,\ref{sec:theory-sec} and \ref{sec:np-input}. We begin by confronting our predictions to results reported by the STAR collaboration. We then present predictions for ALICE FoCal.

\subsection{Comparison to STAR Di-hadron measurements}

Figure~\ref{fig:star_8_panel} shows the away-side di-$\pi^0$ coincidence probability $CP(\Delta\Phi)$ measured by STAR at forward pseudorapidities ($2.6<\eta<4.0$) in p--p and p--Au collisions at $\sqrt{s_{NN}}=200$~GeV, for several transverse-momentum bin selections. In this analysis we introduce two global parameters: (i) an overall multiplicative $K$ factor controlling the normalization of $CP(\Delta\Phi)$, and (ii) a multiplicative factor $C_{NP}$ controlling the strength of the non-perturbative Sudakov contribution. We determine these parameters by fitting the p--p data in the bin $3<p_{T,\mathrm{trig}}<5$~GeV and $1.5<p_{T,\mathrm{asso}}<2.0$~GeV, which features relatively small experimental uncertainties and comparatively reduced sensitivity to fragmentation-systematics. The preferred values are $K=1.5$ and $C_{NP}=2.3$. These values are then used for all the other predictions in p-p and p-A.

Overall, our CGC-based predictions provide a good description of the di-$\pi^0$ correlations in p--Au collisions across the eight $p_T$ bins. For p--p collisions, the agreement is more mixed, with reasonable description in panels (b), (d), and (e), and larger discrepancies in the remaining panels of Fig.~\ref{fig:star_8_panel}. A plausible origin of these tensions is the limited theoretical control over fragmentation at low hadron transverse momentum: the coincidence probability is particularly sensitive because its numerator involves two fragmentation functions, while the denominator involves only one. In this ratio, fragmentation effects associated with the trigger hadron can partially cancel, whereas those associated with the associated hadron persist. This may help explain why bins sharing the same trigger-$p_T$ range are generally better reproduced.

In this work we use the JAM19 fragmentation functions~\cite{Sato:2019yez}. Using alternative sets, such as NP23~\cite{Gao:2024nkz} or DSS07~\cite{deFlorian:2007aj}, can lead to sizable changes in the overall normalization: for instance, DSS07 yields $K\simeq1.25$, whereas NP23 requires $K\simeq 6$. This illustrates the level of normalization uncertainty associated with fragmentation, especially in the low-$p_T$ region. In addition, the STAR coincidence probability has not been corrected for detector efficiencies~\cite{STAR:2021fgw}, and the measured sample includes contributions from decay products of heavier hadrons, which can further affect the normalization. These normalization systematics can be significantly reduced by focusing on ratios between nuclear and proton systems, such as $CP^{p\mathrm{Au}}(\Delta\Phi)/CP^{pp}(\Delta\Phi)$.

An alternative way to characterize the di-pion correlation is to fit a Gaussian distribution in the angular window $0.7\pi < \Delta\Phi < 1.3\pi$ after subtraction of the pedestal. The STAR collaboration has performed this subtraction assuming a constant pedestal. One can then calculate the width and the area (under the window $[0.7\pi, 1.3\pi]$) of the Gaussian. The ratio of width in p-Au to p-p is then a measure of the broadening; similarly, the ratio of areas in p-Au to p-p is a measure of the suppression.  Following \cite{STAR:2021fgw}, the {\it width} is defined as the Gaussian width parameter obtained from a Gaussian + constant fit performed over the angular window $[0.7\pi,\,1.3\pi]$. The {\it area} is defined as the integral of $CP(\Delta\Phi)$ over the same window, evaluated for both the data and the theory curves. Figure~\ref{fig:ratio_com_star} shows the ratios of the area and width of di-hadron correlations in p–Au relative to p–p collisions at 200 GeV. The left panels display these ratios as a function of the associated hadron $p_T$, while the right panels present them as a function of the trigger hadron $p_T$. The area of the forward di-pion coincidence probability in p–Au collisions is approximately 75\% of that in p–p collisions, which can be attributed to larger non-linear evolution effects in Au compared to p. Moreover, the widths of the di-hadron correlations are very similar in both systems, suggesting that the width is predominantly governed by the Sudakov factor (or, equivalently, by parton shower dynamics). The results show that our predictions are consistent with the data within uncertainties. Furthermore, Figure~\ref{fig:ratio_ff_com_star} shows that using different fragmentation functions does not affect these findings. Although the overall normalizations carry significant uncertainties due to the choice of fragmentation functions, it is remarkable that the ratio presented in Figure~\ref{fig:ratio_ff_com_star} remains independent of this choice.

The validity of the dilute-dense approximation is justified by computing the average values of $x_p$ of the proton and $x_g$ of the nucleus probed in each bin. Indeed, at RHIC energies, and for the kinematics examined by the STAR collaboration, $\langle x_g \rangle$ is below $10^{-3}$ (Figure~\ref{fig:xp} right panel), while $x_p \sim 0.5-0.8$ (Figure~\ref{fig:xp} left panel). 

 The Sudakov resummation is parametrically required in the back-to-back regime, where the transverse momentum imbalance
$k_T \equiv |\ptone + \pttwo|$
is much smaller than the hard scale $Q^2= x_p x_g s$, such that
$\alpha_s \ln^2(Q^2/k_T^2) \gtrsim \mathcal{O}(1)$.
This corresponds to
$Q^2/k_T^2 \gtrsim \exp\!\left(1/\sqrt{\alpha_s}\right)$.
Using a representative value $\alpha_s\simeq 0.3$ gives the estimate
$Q^2/k_T^2 \gtrsim 6$. Fig.~\ref{fig:ratio_Q2_kT} in the appendix shows $\langle Q^2/k^2_T\rangle$ for representative cases: the highest-$p_T$ bin, the most asymmetric bin, and the smallest-$p_T$ bin. It shows that  within $0.7\pi<\Delta\phi<1.3\pi$ the ratio $\langle Q^2/k_T^2\rangle$ as a function of $\Delta\phi$ is typically of order (and often above) the above threshold in the relevant kinematic bins, indicating that the cross section in this window is still dominated by configurations with sizable Sudakov double logarithms. 

\subsection{Predictions for FoCal}
We next predict the di-$\pi^0$ correlations anticipated for FoCal kinematics at $\sqrt{s_{NN}} = 5.36$ TeV. Figure \ref{fig:FoCal} displays the di-pion correlations in p-p and p-Pb collisions at FoCal for the forward rapidity range $3.2 < \eta < 5.8$. In this analysis, we include a global rapidity shift of $\Delta y = 0.465$ in the proton-going direction to account for the offset between the center-of-mass and laboratory frames in p-Pb collisions at the LHC. As seen in the figure, FoCal exhibits considerably less suppression in the di-$\pi^0$ correlations (when comparing p-Pb to p-p) than that observed in the STAR measurements, even though FoCal probes a much smaller gluon energy fraction (as demonstrated in Figure \ref{fig:xp} right panel). A more transparent difference between the systems is revealed by the relative area ratio between p-A (Au or Pb) and p-p shown in the panels (a) and (c) in Figure \ref{fig:ratio_com}. Furthermore, we have verified that this conclusion remains unchanged when using alternative fragmentation functions (see panels (a) and (c) in Figure \ref{fig:ratio_ff_com}).

A detailed study shows that the observed behavior arises from the different values of $\langle k_\perp/Q_s \rangle$ probed by STAR and FoCal, where $k_\perp$ is the transverse momentum of the small-$x$ gluon in the nucleus. Here the saturation scale is defined  by $Q^2_s(x)=2/R^2_{\perp,s}(x)$, with $R_{\perp,s}(x)$ chosen so that $\calN(x, R_{\perp,s}(x)) = 1-{\rm exp(-1/2)}$. Figure \ref{fig:kT_Qs} shows that STAR probes a region where $k_\perp/Q_s$ is approximately in the range $1\sim 1.5$, while the FoCal measurements in $3.2 < \eta < 5.8$ at $\sqrt{s_{NN}}=5.36 $ TeV correspond to values around $5\sim6$. Since larger values of $k_\perp/Q_s$ indicate that the measurement is probing a more dilute gluon density, where saturation effects are diminished, FoCal naturally observes smaller saturation effects compared to STAR. This difference is rooted in the distinct fragmentation energy fractions at RHIC and LHC energies. Figure \ref{fig:xp} right panel further illustrates this point by showing the values of $\langle z\rangle$ at STAR and FoCal. Notably, FoCal has a smaller value than STAR, reflecting its higher collision energy ($5.36$ TeV compared to 200 GeV at STAR), which provides a larger phase space for the fragmentation process and thus probes a smaller $z$ region. This increased phase space leads to the larger $\langle k_\perp/Q_s \rangle$ at FoCal and consequently to weaker saturation effects, despite the larger saturation scales probed.

Moreover, the suppression ratio exhibits only a slight dependence on the fragmentation function. For example, Figure \ref{fig:ratio_ff_com} panel (a) shows that using the NP-23 fragmentation function yields slightly stronger suppression than when considering the full rapidity window, yet the suppression remains considerably weaker than that observed at STAR. This consistency indicates that the result is robust and does not crucially depend on the fragmentation function.

A detailed study (right panel of Figure \ref{fig:kT_Qs}) reveals that the average ratio \(\langle k_\perp / Q_s \rangle\) in the FoCal region is somewhat lower than that obtained over the full rapidity range; however, its value is still around 3.5. This relatively large value suggests that even in this most forward region, the probe continues to access the dilute gluon density regime. Similar features can also be observed for ratio between pAu(Pb) relative to pp  as a function of trigger $p_T$, given the associate transverse momentum bin, $1.5<p_{\rm T, asso.}< 2.0$ GeV, shown in the  panels (c)  in Figure~\ref{fig:ratio_com}  and Figure~\ref{fig:ratio_ff_com}.

We further examined the relative width of the di-hadron correlations for both p-A and p-p collisions across STAR and FoCal energies. As shown in the panels (b) and (d) of Figure \ref{fig:ratio_com}, our results indicate that the correlation width is primarily determined by the Sudakov factor (or equivalently, by parton shower dynamics) and Figure \ref{fig:ratio_ff_com} panels (b) and (d) show that this ratio is largely insensitive to the choice of fragmentation functions. Consequently, the width ratio between p-A and p-p remains close to unity at both RHIC and LHC energies. This conclusion is consistent with earlier findings in Ref. \cite{Perepelitsa:2025qpz,Cassar:2025vdp}.  We have additionally checked the calculations for p--p and p--Pb collisions at $\sqrt{s}=5.36$TeV in the interval $2.5<\eta<3.0$. The resulting area ratio and $\langle z\rangle$ follow the same qualitative trend expected in the CGC framework.

\section{Summary and Outlook}
\label{sec:conclusions}
Our analysis demonstrates that the Color Glass Condensate formalism (within the improved TMD approximation) provides a reasonable description of the forward di-$\pi^0$ production around the away-side peak measured by STAR in both proton–proton and proton–gold collisions. While we demonstrate that the coincidence probability $CP(\Delta\Phi)$ is largely sensitive to hadron fragmentation effects, we show that the ratios of the integrated area and the width of the correlation are robust and largely unaffected by the choice of fragmentation function. This insensitivity suggests that these ratios serve as reliable probes of the underlying initial-state dynamics; thus, providing a window to search for gluon saturation effects.

We also predict these observables at kinematics accessible at the ALICE FoCal experiment at $\sqrt{s_{NN}} = 5.36\,\text{TeV}$.  We predict significantly less suppression in di-hadron correlations compared to those measured at STAR. Despite the larger saturation scales probed at FoCal, thanks to the smaller-$x$, due to larger center of mass energy and more forward rapidities, the average values of fragmentation energy $\langle z \rangle$ are smaller, which in turn implies that partons tend to have significantly larger transverse momentum; hence, are more sensitive to the dilute regime.

To further enhance sensitivity to saturation effects in FoCal measurements, complementary observables, such as energy correlators \cite{Kang:2025vjk,Ganguli:2025aqa,Moult:2025nhu}, heavy-flavor production \cite{Marquet:2017xwy,Altinoluk:2021ygv,Marquet:2025jdr}, or $\gamma-$jet \cite{Ganguli:2023joy} and di-jet correlations \cite{Al-Mashad:2022zbq} could be employed. These additional measurements provide a more direct access to the kinematics of the underlying partonic process; thus amplifying the saturation signals.

Another possibility is to examine the dependence of the suppression and broadening of the coincidence probability as a function of impact parameter of the collision (see e.g. \cite{Giacalone:2018fbc}). The ALICE and LHCb collaborations have reported some measurements in this direction \cite{ALICE:2012eyl,LHCb:2015coe}. We expect stronger gluon saturation effects in more central collisions as the density of gluons is generally larger. However, in experiments, centrality is typically determined via multiplicity, which does not always correlate strongly with the impact parameter in p-p and p-A collisions. Looking ahead, if a method can be found to better constrain the impact parameter, a promising approach is to construct ratio observables relative to minimum-bias events. These can minimize centrality-independent contributions, such as Sudakov effects, thereby isolating the saturation-driven suppression.

In addition to these complementary studies, we hope that in the near future full next-to-leading order (NLO) contributions will be available for di-hadron production. Some partial efforts in this direction can be found in \cite{Iancu:2022gpw,Caucal:2025xxh}, which, coupled to existing next-to-leading order computations for single inclusive hadron production \cite{Chirilli:2011km,Shi:2021hwx,Wang:2022zdu}, can provide a path forward to full NLO precision and a better assessment of theoretical uncertainties.

Finally, it is also important to note that alternative models based on nuclear parton distribution functions (nPDFs) \cite{Perepelitsa:2025qpz} or dynamical shadowing and cold nuclear matter energy loss \cite{Kang:2011bp} can both qualitatively reproduce the di-hadron data and offer complementary predictions on energy dependence. 
Systematic comparisons among CGC-based calculations, nPDF frameworks, and cold nuclear matter energy loss models, together with efforts to elucidate the connections between these different formalisms~\cite{Fu:2023jqv,Fu:2024sba}, may help disentangle genuine gluon saturation effects from cold and hot nuclear matter modifications~\cite{Arleo:2025oos}.

\begin{acknowledgements}
We thank X. Chu, C. Marquet, H. Xing and  F. Yuan for providing helpful discussions. Z.K., F.S., and W.Z. are grateful to Kaiyuan Shi for collaboration at early stages of this project. This work is supported by the U.S. Department of Energy, Office of Science, Office of Nuclear Physics, under DOE Contract No.~DE-SC0012704 (B.P.S.), and within the framework of the Saturated Glue (SURGE) Topical Theory Collaboration (Z.K., F.S., B.P.S., R.V., W.Z.). PC is funded by the Agence Nationale de la Recherche under grant ANR-25-CE31-5230 (TMD-SAT). Z.K was also supported by the National Science Foundation under Grants No.~PHY-2515057.
P.K. acknowledges the support of the Polish National Science Center (NCN) grant No.~2022/46/E/ST2/00346. He gratefully acknowledges Polish high-performance computing infrastructure PLGrid (HPC Center: ACK Cyfronet AGH) for providing computer facilities and support within computational grant no. PLG/2024/017690. F.S. is supported by the Laboratory Directed Research and Development of Brookhaven National Laboratory and RIKEN-BNL Research Center. 
T.S. acknowledges the support of the Polish National Science Center (NCN) Grant No. 2021/43/D/ST2/03375.
R.V. was  supported by U.S. Department of Energy, Office of Science,
under contract DE-SC0012704. He was also supported at Stony Brook by the Simons Foundation as a co-PI under Award number 994318 (Simons Collaboration on Confinement and QCD Strings). He acknowledges support from the Royal Society Wolfson Foundation Visiting Fellowship and the hospitality of the Higgs Center at the University of Edinburgh. 
W.Z. was supported in part by the National Science Foundation (NSF) within the framework of the JETSCAPE collaboration (OAC-2004571). F.S. and W.Z. also thank the Institute for Nuclear Theory where part of this work was carried out during the ``Precision QCD with the Electron Ion Collider" program, and the participants of the program for stimulating discussions.

The content of this article does not reflect the official opinion of the European Union, and responsibility for the information and views expressed therein lies entirely with the authors.
\end{acknowledgements}

\bibliographystyle{utcaps}
\bibliography{refs}

\providecommand{\href}[2]{#2}\begingroup\raggedright\begin{thebibliography}{100}

\bibitem{Gribov:1983ivg}
L.~V. Gribov, E.~M. Levin, and M.~G. Ryskin, ``{Semihard Processes in QCD},'' \href{http://dx.doi.org/10.1016/0370-1573(83)90022-4}{{\em Phys. Rept.} {\bfseries 100} (1983) 1--150}.

\bibitem{Mueller:1985wy}
A.~H. Mueller and J.-w. Qiu, ``{Gluon Recombination and Shadowing at Small Values of x},'' \href{http://dx.doi.org/10.1016/0550-3213(86)90164-1}{{\em Nucl. Phys. B} {\bfseries 268} (1986) 427--452}.

\bibitem{McLerran:1993ni}
L.~D. McLerran and R.~Venugopalan, ``{Computing quark and gluon distribution functions for very large nuclei},'' \href{http://dx.doi.org/10.1103/PhysRevD.49.2233}{{\em Phys. Rev. D} {\bfseries 49} (1994) 2233--2241}, \href{http://arxiv.org/abs/hep-ph/9309289}{{\ttfamily arXiv:hep-ph/9309289}}.

\bibitem{McLerran:1993ka}
L.~D. McLerran and R.~Venugopalan, ``{Gluon distribution functions for very large nuclei at small transverse momentum},'' \href{http://dx.doi.org/10.1103/PhysRevD.49.3352}{{\em Phys. Rev. D} {\bfseries 49} (1994) 3352--3355}, \href{http://arxiv.org/abs/hep-ph/9311205}{{\ttfamily arXiv:hep-ph/9311205}}.

\bibitem{McLerran:1994vd}
L.~D. McLerran and R.~Venugopalan, ``{Green's functions in the color field of a large nucleus},'' \href{http://dx.doi.org/10.1103/PhysRevD.50.2225}{{\em Phys. Rev. D} {\bfseries 50} (1994) 2225--2233}, \href{http://arxiv.org/abs/hep-ph/9402335}{{\ttfamily arXiv:hep-ph/9402335}}.

\bibitem{Ayala:1995kg}
A.~Ayala, J.~Jalilian-Marian, L.~D. McLerran, and R.~Venugopalan, ``{The Gluon propagator in nonAbelian Weizsacker-Williams fields},'' \href{http://dx.doi.org/10.1103/PhysRevD.52.2935}{{\em Phys. Rev. D} {\bfseries 52} (1995) 2935--2943}, \href{http://arxiv.org/abs/hep-ph/9501324}{{\ttfamily arXiv:hep-ph/9501324}}.

\bibitem{Ayala:1995hx}
A.~Ayala, J.~Jalilian-Marian, L.~D. McLerran, and R.~Venugopalan, ``{Quantum corrections to the Weizsacker-Williams gluon distribution function at small x},'' \href{http://dx.doi.org/10.1103/PhysRevD.53.458}{{\em Phys. Rev. D} {\bfseries 53} (1996) 458--475}, \href{http://arxiv.org/abs/hep-ph/9508302}{{\ttfamily arXiv:hep-ph/9508302}}.

\bibitem{Iancu:2003xm}
E.~Iancu and R.~Venugopalan, {\em {The Color glass condensate and high-energy scattering in QCD}}, \href{http://dx.doi.org/10.1142/9789812795533_0005}{pp.~249--3363}.
\newblock 3, 2003.
\newblock \href{http://arxiv.org/abs/hep-ph/0303204}{{\ttfamily arXiv:hep-ph/0303204}}.

\bibitem{Gelis:2010nm}
F.~Gelis, E.~Iancu, J.~Jalilian-Marian, and R.~Venugopalan, ``{The Color Glass Condensate},'' \href{http://dx.doi.org/10.1146/annurev.nucl.010909.083629}{{\em Ann. Rev. Nucl. Part. Sci.} {\bfseries 60} (2010) 463--489}, \href{http://arxiv.org/abs/1002.0333}{{\ttfamily arXiv:1002.0333 [hep-ph]}}.

\bibitem{Kovchegov:2012mbw}
Y.~V. Kovchegov and E.~Levin, \href{http://dx.doi.org/10.1017/9781009291446}{{\em {Quantum Chromodynamics at High Energy}}}, vol.~33.
\newblock Oxford University Press, 2013.

\bibitem{Albacete:2014fwa}
J.~L. Albacete and C.~Marquet, ``{Gluon saturation and initial conditions for relativistic heavy ion collisions},'' \href{http://dx.doi.org/10.1016/j.ppnp.2014.01.004}{{\em Prog. Part. Nucl. Phys.} {\bfseries 76} (2014) 1--42}, \href{http://arxiv.org/abs/1401.4866}{{\ttfamily arXiv:1401.4866 [hep-ph]}}.

\bibitem{Morreale:2021pnn}
A.~Morreale and F.~Salazar, ``{Mining for Gluon Saturation at Colliders},'' \href{http://dx.doi.org/10.3390/universe7080312}{{\em Universe} {\bfseries 7} no.~8, (2021) 312}, \href{http://arxiv.org/abs/2108.08254}{{\ttfamily arXiv:2108.08254 [hep-ph]}}.

\bibitem{JalilianMarian:1996xn}
J.~Jalilian-Marian, A.~Kovner, L.~D. McLerran, and H.~Weigert, ``{The Intrinsic glue distribution at very small x},'' \href{http://dx.doi.org/10.1103/PhysRevD.55.5414}{{\em Phys. Rev. D} {\bfseries 55} (1997) 5414--5428}, \href{http://arxiv.org/abs/hep-ph/9606337}{{\ttfamily arXiv:hep-ph/9606337}}.

\bibitem{JalilianMarian:1997dw}
J.~Jalilian-Marian, A.~Kovner, and H.~Weigert, ``{The Wilson renormalization group for low x physics: Gluon evolution at finite parton density},'' \href{http://dx.doi.org/10.1103/PhysRevD.59.014015}{{\em Phys. Rev.} {\bfseries D59} (1998) 014015},
\href{http://arxiv.org/abs/hep-ph/9709432}{{\ttfamily arXiv:hep-ph/9709432 [hep-ph]}}.

\bibitem{Kovner:2000pt}
A.~Kovner, J.~G. Milhano, and H.~Weigert, ``{Relating different approaches to nonlinear QCD evolution at finite gluon density},'' \href{http://dx.doi.org/10.1103/PhysRevD.62.114005}{{\em Phys. Rev. D} {\bfseries 62} (2000) 114005}, \href{http://arxiv.org/abs/hep-ph/0004014}{{\ttfamily arXiv:hep-ph/0004014}}.

\bibitem{Iancu:2000hn}
E.~Iancu, A.~Leonidov, and L.~D. McLerran, ``{Nonlinear gluon evolution in the color glass condensate. 1.},'' \href{http://dx.doi.org/10.1016/S0375-9474(01)00642-X}{{\em Nucl. Phys.} {\bfseries A692} (2001) 583--645},
\href{http://arxiv.org/abs/hep-ph/0011241}{{\ttfamily arXiv:hep-ph/0011241 [hep-ph]}}.

\bibitem{Iancu:2001ad}
E.~Iancu, A.~Leonidov, and L.~D. McLerran, ``{The Renormalization group equation for the color glass condensate},'' \href{http://dx.doi.org/10.1016/S0370-2693(01)00524-X}{{\em Phys. Lett.} {\bfseries B510} (2001) 133--144},
\href{http://arxiv.org/abs/hep-ph/0102009}{{\ttfamily arXiv:hep-ph/0102009 [hep-ph]}}.

\bibitem{Ferreiro:2001qy}
E.~Ferreiro, E.~Iancu, A.~Leonidov, and L.~McLerran, ``{Nonlinear gluon evolution in the color glass condensate. 2.},'' \href{http://dx.doi.org/10.1016/S0375-9474(01)01329-X}{{\em Nucl. Phys. A} {\bfseries 703} (2002) 489--538}, \href{http://arxiv.org/abs/hep-ph/0109115}{{\ttfamily arXiv:hep-ph/0109115}}.

\bibitem{Weigert:2005us}
H.~Weigert, ``{Evolution at small x(bj): The Color glass condensate},'' \href{http://dx.doi.org/10.1016/j.ppnp.2005.01.029}{{\em Prog. Part. Nucl. Phys.} {\bfseries 55} (2005) 461--565}, \href{http://arxiv.org/abs/hep-ph/0501087}{{\ttfamily arXiv:hep-ph/0501087}}.

\bibitem{Balitsky:1995ub}
I.~Balitsky, ``{Operator expansion for high-energy scattering},'' \href{http://dx.doi.org/10.1016/0550-3213(95)00638-9}{{\em Nucl. Phys.} {\bfseries B463} (1996) 99--160},
\href{http://arxiv.org/abs/hep-ph/9509348}{{\ttfamily arXiv:hep-ph/9509348 [hep-ph]}}.

\bibitem{Kovchegov:1999yj}
Y.~V. Kovchegov, ``{Small x F(2) structure function of a nucleus including multiple pomeron exchanges},'' \href{http://dx.doi.org/10.1103/PhysRevD.60.034008}{{\em Phys. Rev.} {\bfseries D60} (1999) 034008},
\href{http://arxiv.org/abs/hep-ph/9901281}{{\ttfamily arXiv:hep-ph/9901281 [hep-ph]}}.

\bibitem{Kharzeev:2004bw}
D.~Kharzeev, E.~Levin, and L.~McLerran, ``{Jet azimuthal correlations and parton saturation in the color glass condensate},'' \href{http://dx.doi.org/10.1016/j.nuclphysa.2004.10.031}{{\em Nucl. Phys. A} {\bfseries 748} (2005) 627--640}, \href{http://arxiv.org/abs/hep-ph/0403271}{{\ttfamily arXiv:hep-ph/0403271}}.

\bibitem{Marquet:2007vb}
C.~Marquet, ``{Forward inclusive dijet production and azimuthal correlations in p(A) collisions},'' \href{http://dx.doi.org/10.1016/j.nuclphysa.2007.09.001}{{\em Nucl. Phys. A} {\bfseries 796} (2007) 41--60}, \href{http://arxiv.org/abs/0708.0231}{{\ttfamily arXiv:0708.0231 [hep-ph]}}.

\bibitem{Albacete:2010pg}
J.~L. Albacete and C.~Marquet, ``{Azimuthal correlations of forward di-hadrons in d+Au collisions at RHIC in the Color Glass Condensate},'' \href{http://dx.doi.org/10.1103/PhysRevLett.105.162301}{{\em Phys. Rev. Lett.} {\bfseries 105} (2010) 162301}, \href{http://arxiv.org/abs/1005.4065}{{\ttfamily arXiv:1005.4065 [hep-ph]}}.

\bibitem{Kharzeev:2003wz}
D.~Kharzeev, Y.~V. Kovchegov, and K.~Tuchin, ``{Cronin effect and high p(T) suppression in pA collisions},'' \href{http://dx.doi.org/10.1103/PhysRevD.68.094013}{{\em Phys. Rev. D} {\bfseries 68} (2003) 094013}, \href{http://arxiv.org/abs/hep-ph/0307037}{{\ttfamily arXiv:hep-ph/0307037}}.

\bibitem{Albacete:2003iq}
J.~L. Albacete, N.~Armesto, A.~Kovner, C.~A. Salgado, and U.~A. Wiedemann, ``{Energy dependence of the Cronin effect from nonlinear QCD evolution},'' \href{http://dx.doi.org/10.1103/PhysRevLett.92.082001}{{\em Phys. Rev. Lett.} {\bfseries 92} (2004) 082001}, \href{http://arxiv.org/abs/hep-ph/0307179}{{\ttfamily arXiv:hep-ph/0307179}}.

\bibitem{Lappi:2012nh}
T.~Lappi and H.~Mantysaari, ``{Forward dihadron correlations in deuteron-gold collisions with the Gaussian approximation of JIMWLK},'' \href{http://dx.doi.org/10.1016/j.nuclphysa.2013.03.017}{{\em Nucl. Phys. A} {\bfseries 908} (2013) 51--72}, \href{http://arxiv.org/abs/1209.2853}{{\ttfamily arXiv:1209.2853 [hep-ph]}}.

\bibitem{Dominguez:2011wm}
F.~Dominguez, C.~Marquet, B.-W. Xiao, and F.~Yuan, ``{Universality of Unintegrated Gluon Distributions at small x},'' \href{http://dx.doi.org/10.1103/PhysRevD.83.105005}{{\em Phys. Rev. D} {\bfseries 83} (2011) 105005}, \href{http://arxiv.org/abs/1101.0715}{{\ttfamily arXiv:1101.0715 [hep-ph]}}.

\bibitem{Dominguez:2010xd}
F.~Dominguez, B.-W. Xiao, and F.~Yuan, ``{$k_t$-factorization for Hard Processes in Nuclei},'' \href{http://dx.doi.org/10.1103/PhysRevLett.106.022301}{{\em Phys. Rev. Lett.} {\bfseries 106} (2011) 022301}, \href{http://arxiv.org/abs/1009.2141}{{\ttfamily arXiv:1009.2141 [hep-ph]}}.

\bibitem{Marquet:2016cgx}
C.~Marquet, E.~Petreska, and C.~Roiesnel, ``{Transverse-momentum-dependent gluon distributions from JIMWLK evolution},'' \href{http://dx.doi.org/10.1007/JHEP10(2016)065}{{\em JHEP} {\bfseries 10} (2016) 065}, \href{http://arxiv.org/abs/1608.02577}{{\ttfamily arXiv:1608.02577 [hep-ph]}}.

\bibitem{Petreska:2018cbf}
E.~Petreska, ``{TMD gluon distributions at small x in the CGC theory},'' \href{http://dx.doi.org/10.1142/S0218301318300035}{{\em Int. J. Mod. Phys. E} {\bfseries 27} no.~05, (2018) 1830003}, \href{http://arxiv.org/abs/1804.04981}{{\ttfamily arXiv:1804.04981 [hep-ph]}}.

\bibitem{Cali:2021tsh}
S.~Cali, K.~Cichy, P.~Korcyl, P.~Kotko, K.~Kutak, and C.~Marquet, ``{On systematic effects in the numerical solutions of the JIMWLK equation},'' \href{http://dx.doi.org/10.1140/epjc/s10052-021-09380-6}{{\em Eur. Phys. J. C} {\bfseries 81} no.~7, (2021) 663}, \href{http://arxiv.org/abs/2104.14254}{{\ttfamily arXiv:2104.14254 [hep-ph]}}.

\bibitem{Bomhof:2006dp}
C.~J. Bomhof, P.~J. Mulders, and F.~Pijlman, ``{The Construction of gauge-links in arbitrary hard processes},'' \href{http://dx.doi.org/10.1140/epjc/s2006-02554-2}{{\em Eur. Phys. J. C} {\bfseries 47} (2006) 147--162}, \href{http://arxiv.org/abs/hep-ph/0601171}{{\ttfamily arXiv:hep-ph/0601171}}.

\bibitem{Dominguez:2012ad}
F.~Dominguez, C.~Marquet, A.~M. Stasto, and B.-W. Xiao, ``{Universality of multiparticle production in QCD at high energies},'' \href{http://dx.doi.org/10.1103/PhysRevD.87.034007}{{\em Phys. Rev. D} {\bfseries 87} (2013) 034007}, \href{http://arxiv.org/abs/1210.1141}{{\ttfamily arXiv:1210.1141 [hep-ph]}}.

\bibitem{Jalilian-Marian:2012wwi}
J.~Jalilian-Marian and A.~H. Rezaeian, ``{Prompt photon production and photon-hadron correlations at RHIC and the LHC from the Color Glass Condensate},'' \href{http://dx.doi.org/10.1103/PhysRevD.86.034016}{{\em Phys. Rev. D} {\bfseries 86} (2012) 034016}, \href{http://arxiv.org/abs/1204.1319}{{\ttfamily arXiv:1204.1319 [hep-ph]}}.

\bibitem{Stasto:2011ru}
A.~Stasto, B.-W. Xiao, and F.~Yuan, ``{Back-to-Back Correlations of Di-hadrons in dAu Collisions at RHIC},'' \href{http://dx.doi.org/10.1016/j.physletb.2012.08.044}{{\em Phys. Lett. B} {\bfseries 716} (2012) 430--434}, \href{http://arxiv.org/abs/1109.1817}{{\ttfamily arXiv:1109.1817 [hep-ph]}}.

\bibitem{Kotko:2017oxg}
P.~Kotko, K.~Kutak, S.~Sapeta, A.~M. Stasto, and M.~Strikman, ``{Estimating nonlinear effects in forward dijet production in ultra-peripheral heavy ion collisions at the LHC},'' \href{http://dx.doi.org/10.1140/epjc/s10052-017-4906-6}{{\em Eur. Phys. J. C} {\bfseries 77} no.~5, (2017) 353}, \href{http://arxiv.org/abs/1702.03063}{{\ttfamily arXiv:1702.03063 [hep-ph]}}.

\bibitem{Albacete:2018ruq}
J.~L. Albacete, G.~Giacalone, C.~Marquet, and M.~Matas, ``{Forward dihadron back-to-back correlations in $pA$ collisions},'' \href{http://dx.doi.org/10.1103/PhysRevD.99.014002}{{\em Phys. Rev. D} {\bfseries 99} no.~1, (2019) 014002}, \href{http://arxiv.org/abs/1805.05711}{{\ttfamily arXiv:1805.05711 [hep-ph]}}.

\bibitem{vanHameren:2019ysa}
A.~van Hameren, P.~Kotko, K.~Kutak, and S.~Sapeta, ``{Broadening and saturation effects in dijet azimuthal correlations in p-p and p-Pb collisions at $\mathbf{\sqrt{s}} = $ 5.02 TeV},'' \href{http://dx.doi.org/10.1016/j.physletb.2019.06.055}{{\em Phys. Lett. B} {\bfseries 795} (2019) 511--515}, \href{http://arxiv.org/abs/1903.01361}{{\ttfamily arXiv:1903.01361 [hep-ph]}}.

\bibitem{vanHameren:2020rqt}
A.~van Hameren, P.~Kotko, K.~Kutak, and S.~Sapeta, ``{Sudakov effects in central-forward dijet production in high energy factorization},'' \href{http://dx.doi.org/10.1016/j.physletb.2021.136078}{{\em Phys. Lett. B} {\bfseries 814} (2021) 136078}, \href{http://arxiv.org/abs/2010.13066}{{\ttfamily arXiv:2010.13066 [hep-ph]}}.

\bibitem{Benic:2022ixp}
S.~Beni\'c, O.~Garcia-Montero, and A.~Perkov, ``{Isolated photon-hadron production in high energy pp and pA collisions at RHIC and LHC},'' \href{http://dx.doi.org/10.1103/PhysRevD.105.114052}{{\em Phys. Rev. D} {\bfseries 105} no.~11, (2022) 114052}, \href{http://arxiv.org/abs/2203.01685}{{\ttfamily arXiv:2203.01685 [hep-ph]}}.

\bibitem{Al-Mashad:2022zbq}
M.~A. Al-Mashad, A.~van Hameren, H.~Kakkad, P.~Kotko, K.~Kutak, P.~van Mechelen, and S.~Sapeta, ``{Dijet azimuthal correlations in p-p and p-Pb collisions at forward LHC calorimeters},'' \href{http://dx.doi.org/10.1007/JHEP12(2022)131}{{\em JHEP} {\bfseries 12} (2022) 131}, \href{http://arxiv.org/abs/2210.06613}{{\ttfamily arXiv:2210.06613 [hep-ph]}}.

\bibitem{vanHameren:2023oiq}
A.~van Hameren, H.~Kakkad, P.~Kotko, K.~Kutak, and S.~Sapeta, ``{Searching for saturation in forward dijet production at the LHC},'' \href{http://dx.doi.org/10.1140/epjc/s10052-023-12120-7}{{\em Eur. Phys. J. C} {\bfseries 83} no.~10, (2023) 947}, \href{http://arxiv.org/abs/2306.17513}{{\ttfamily arXiv:2306.17513 [hep-ph]}}.

\bibitem{Ganguli:2023joy}
I.~Ganguli, A.~van Hameren, P.~Kotko, and K.~Kutak, ``{Forward $\gamma $+jet production in proton-proton and proton-lead collisions at LHC within the FoCal calorimeter acceptance},'' \href{http://dx.doi.org/10.1140/epjc/s10052-023-12043-3}{{\em Eur. Phys. J. C} {\bfseries 83} no.~9, (2023) 868}, \href{http://arxiv.org/abs/2306.04706}{{\ttfamily arXiv:2306.04706 [hep-ph]}}.

\bibitem{Zheng:2014vka}
L.~Zheng, E.~C. Aschenauer, J.~H. Lee, and B.-W. Xiao, ``{Probing Gluon Saturation through Dihadron Correlations at an Electron-Ion Collider},'' \href{http://dx.doi.org/10.1103/PhysRevD.89.074037}{{\em Phys. Rev. D} {\bfseries 89} no.~7, (2014) 074037}, \href{http://arxiv.org/abs/1403.2413}{{\ttfamily arXiv:1403.2413 [hep-ph]}}.

\bibitem{vanHameren:2021sqc}
A.~van Hameren, P.~Kotko, K.~Kutak, S.~Sapeta, and E.~\.Zar\'ow, ``{Probing gluon number density with electron-dijet correlations at EIC},'' \href{http://dx.doi.org/10.1140/epjc/s10052-021-09529-3}{{\em Eur. Phys. J. C} {\bfseries 81} no.~8, (2021) 741}, \href{http://arxiv.org/abs/2106.13964}{{\ttfamily arXiv:2106.13964 [hep-ph]}}.

\bibitem{Braidot:2010zh}
{\bfseries STAR} Collaboration, E.~Braidot, ``{Suppression of Forward Pion Correlations in d+Au Interactions at STAR},'' in {\em {45th Rencontres de Moriond on QCD and High Energy Interactions}}, pp.~355--338.
\newblock 5, 2010.
\newblock \href{http://arxiv.org/abs/1005.2378}{{\ttfamily arXiv:1005.2378 [hep-ph]}}.

\bibitem{STAR:2021fgw}
{\bfseries STAR} Collaboration, M.~S. Abdallah {\em et al.}, ``{Evidence for Nonlinear Gluon Effects in QCD and Their Mass Number Dependence at STAR},'' \href{http://dx.doi.org/10.1103/PhysRevLett.129.092501}{{\em Phys. Rev. Lett.} {\bfseries 129} no.~9, (2022) 092501}, \href{http://arxiv.org/abs/2111.10396}{{\ttfamily arXiv:2111.10396 [nucl-ex]}}.

\bibitem{PHENIX:2011puq}
{\bfseries PHENIX} Collaboration, A.~Adare {\em et al.}, ``{Suppression of back-to-back hadron pairs at forward rapidity in $d+$Au Collisions at $\sqrt{s_{NN}}=200$ GeV},'' \href{http://dx.doi.org/10.1103/PhysRevLett.107.172301}{{\em Phys. Rev. Lett.} {\bfseries 107} (2011) 172301}, \href{http://arxiv.org/abs/1105.5112}{{\ttfamily arXiv:1105.5112 [nucl-ex]}}.

\bibitem{ATLAS:2019jgo}
{\bfseries ATLAS} Collaboration, M.~Aaboud {\em et al.}, ``{Dijet azimuthal correlations and conditional yields in pp and p+Pb collisions at sNN=5.02TeV with the ATLAS detector},'' \href{http://dx.doi.org/10.1103/PhysRevC.100.034903}{{\em Phys. Rev. C} {\bfseries 100} no.~3, (2019) 034903}, \href{http://arxiv.org/abs/1901.10440}{{\ttfamily arXiv:1901.10440 [nucl-ex]}}.

\bibitem{Perepelitsa:2025qpz}
D.~V. Perepelitsa, ``{Description of di-hadron saturation signals within a universal nuclear parton distribution function approach},'' \href{http://dx.doi.org/10.1103/PhysRevC.111.054901}{{\em Phys. Rev. C} {\bfseries 111} no.~5, (2025) 054901}, \href{http://arxiv.org/abs/2501.18347}{{\ttfamily arXiv:2501.18347 [nucl-th]}}.

\bibitem{Cassar:2025vdp}
K.~Cassar, Z.~Wang, X.~Chu, and E.-C. Aschenauer, ``{Investigating the broadening phenomenon in two-particle correlations induced by gluon saturation},'' \href{http://dx.doi.org/10.1103/7jxl-8pzn}{{\em Phys. Rev. D} {\bfseries 112} no.~3, (2025) 034034}, \href{http://arxiv.org/abs/2503.08447}{{\ttfamily arXiv:2503.08447 [hep-ph]}}.

\bibitem{Collins:1981uk}
J.~C. Collins and D.~E. Soper, ``{Back-To-Back Jets in QCD},'' \href{http://dx.doi.org/10.1016/0550-3213(81)90339-4}{{\em Nucl. Phys. B} {\bfseries 193} (1981) 381}. [Erratum: Nucl.Phys.B 213, 545 (1983)].

\bibitem{Collins:1981uw}
J.~C. Collins and D.~E. Soper, ``{Parton Distribution and Decay Functions},'' \href{http://dx.doi.org/10.1016/0550-3213(82)90021-9}{{\em Nucl. Phys. B} {\bfseries 194} (1982) 445--492}.

\bibitem{Collins:1984kg}
J.~C. Collins, D.~E. Soper, and G.~F. Sterman, ``{Transverse Momentum Distribution in Drell-Yan Pair and W and Z Boson Production},'' \href{http://dx.doi.org/10.1016/0550-3213(85)90479-1}{{\em Nucl. Phys. B} {\bfseries 250} (1985) 199--224}.

\bibitem{Collins:2011zzd}
J.~Collins, \href{http://dx.doi.org/10.1017/9781009401845}{{\em {Foundations of Perturbative QCD}}}, vol.~32.
\newblock Cambridge University Press, 2011.

\bibitem{Gao:2023ulg}
M.-S. Gao, Z.-B. Kang, D.~Y. Shao, J.~Terry, and C.~Zhang, ``{QCD resummation of dijet azimuthal decorrelations in pp and pA collisions},'' \href{http://dx.doi.org/10.1007/JHEP10(2023)013}{{\em JHEP} {\bfseries 10} (2023) 013}, \href{http://arxiv.org/abs/2306.09317}{{\ttfamily arXiv:2306.09317 [hep-ph]}}.

\bibitem{Sun:2015doa}
P.~Sun, C.~P. Yuan, and F.~Yuan, ``{Transverse Momentum Resummation for Dijet Correlation in Hadronic Collisions},'' \href{http://dx.doi.org/10.1103/PhysRevD.92.094007}{{\em Phys. Rev. D} {\bfseries 92} no.~9, (2015) 094007}, \href{http://arxiv.org/abs/1506.06170}{{\ttfamily arXiv:1506.06170 [hep-ph]}}.

\bibitem{Kang:2020xez}
Z.-B. Kang, K.~Lee, D.~Y. Shao, and J.~Terry, ``{The Sivers Asymmetry in Hadronic Dijet Production},'' \href{http://dx.doi.org/10.1007/JHEP02(2021)066}{{\em JHEP} {\bfseries 02} (2021) 066}, \href{http://arxiv.org/abs/2008.05470}{{\ttfamily arXiv:2008.05470 [hep-ph]}}.

\bibitem{Mueller:2012uf}
A.~H. Mueller, B.-W. Xiao, and F.~Yuan, ``{Sudakov Resummation in Small-$x$ Saturation Formalism},'' \href{http://dx.doi.org/10.1103/PhysRevLett.110.082301}{{\em Phys. Rev. Lett.} {\bfseries 110} no.~8, (2013) 082301}, \href{http://arxiv.org/abs/1210.5792}{{\ttfamily arXiv:1210.5792 [hep-ph]}}.

\bibitem{Mueller:2013wwa}
A.~H. Mueller, B.-W. Xiao, and F.~Yuan, ``{Sudakov double logarithms resummation in hard processes in the small-x saturation formalism},'' \href{http://dx.doi.org/10.1103/PhysRevD.88.114010}{{\em Phys. Rev. D} {\bfseries 88} no.~11, (2013) 114010}, \href{http://arxiv.org/abs/1308.2993}{{\ttfamily arXiv:1308.2993 [hep-ph]}}.

\bibitem{Zhou:2018lfq}
J.~Zhou, ``{Scale dependence of the small x transverse momentum dependent gluon distribution},'' \href{http://dx.doi.org/10.1103/PhysRevD.99.054026}{{\em Phys. Rev. D} {\bfseries 99} no.~5, (2019) 054026}, \href{http://arxiv.org/abs/1807.00506}{{\ttfamily arXiv:1807.00506 [hep-ph]}}.

\bibitem{Xiao:2017yya}
B.-W. Xiao, F.~Yuan, and J.~Zhou, ``{Transverse Momentum Dependent Parton Distributions at Small-x},'' \href{http://dx.doi.org/10.1016/j.nuclphysb.2017.05.012}{{\em Nucl. Phys. B} {\bfseries 921} (2017) 104--126}, \href{http://arxiv.org/abs/1703.06163}{{\ttfamily arXiv:1703.06163 [hep-ph]}}.

\bibitem{Hatta:2020bgy}
Y.~Hatta, B.-W. Xiao, F.~Yuan, and J.~Zhou, ``{Anisotropy in Dijet Production in Exclusive and Inclusive Processes},'' \href{http://dx.doi.org/10.1103/PhysRevLett.126.142001}{{\em Phys. Rev. Lett.} {\bfseries 126} no.~14, (2021) 142001}, \href{http://arxiv.org/abs/2010.10774}{{\ttfamily arXiv:2010.10774 [hep-ph]}}.

\bibitem{Hatta:2021jcd}
Y.~Hatta, B.-W. Xiao, F.~Yuan, and J.~Zhou, ``{Azimuthal angular asymmetry of soft gluon radiation in jet production},'' \href{http://dx.doi.org/10.1103/PhysRevD.104.054037}{{\em Phys. Rev. D} {\bfseries 104} no.~5, (2021) 054037}, \href{http://arxiv.org/abs/2106.05307}{{\ttfamily arXiv:2106.05307 [hep-ph]}}.

\bibitem{Hentschinski:2021lsh}
M.~Hentschinski, ``{Transverse momentum dependent gluon distribution within high energy factorization at next-to-leading order},'' \href{http://dx.doi.org/10.1103/PhysRevD.104.054014}{{\em Phys. Rev. D} {\bfseries 104} no.~5, (2021) 054014}, \href{http://arxiv.org/abs/2107.06203}{{\ttfamily arXiv:2107.06203 [hep-ph]}}.

\bibitem{Taels:2022tza}
P.~Taels, T.~Altinoluk, G.~Beuf, and C.~Marquet, ``{Dijet photoproduction at low x at next-to-leading order and its back-to-back limit},'' \href{http://dx.doi.org/10.1007/JHEP10(2022)184}{{\em JHEP} {\bfseries 10} (2022) 184}, \href{http://arxiv.org/abs/2204.11650}{{\ttfamily arXiv:2204.11650 [hep-ph]}}.

\bibitem{Caucal:2022ulg}
P.~Caucal, F.~Salazar, B.~Schenke, and R.~Venugopalan, ``{Back-to-back inclusive dijets in DIS at small x: Sudakov suppression and gluon saturation at NLO},'' \href{http://dx.doi.org/10.1007/JHEP11(2022)169}{{\em JHEP} {\bfseries 11} (2022) 169}, \href{http://arxiv.org/abs/2208.13872}{{\ttfamily arXiv:2208.13872 [hep-ph]}}.

\bibitem{Caucal:2023nci}
P.~Caucal, F.~Salazar, B.~Schenke, T.~Stebel, and R.~Venugopalan, ``{Back-to-back inclusive dijets in DIS at small x: gluon Weizs\"acker-Williams distribution at NLO},'' \href{http://dx.doi.org/10.1007/JHEP08(2023)062}{{\em JHEP} {\bfseries 08} (2023) 062}, \href{http://arxiv.org/abs/2304.03304}{{\ttfamily arXiv:2304.03304 [hep-ph]}}.

\bibitem{Caucal:2023fsf}
P.~Caucal, F.~Salazar, B.~Schenke, T.~Stebel, and R.~Venugopalan, ``{Back-to-Back Inclusive Dijets in Deep Inelastic Scattering at Small x: Complete NLO Results and Predictions},'' \href{http://dx.doi.org/10.1103/PhysRevLett.132.081902}{{\em Phys. Rev. Lett.} {\bfseries 132} no.~8, (2024) 081902}, \href{http://arxiv.org/abs/2308.00022}{{\ttfamily arXiv:2308.00022 [hep-ph]}}.

\bibitem{Caucal:2024bae}
P.~Caucal and E.~Iancu, ``{Evolution of the transverse-momentum dependent gluon distribution at small x},'' \href{http://dx.doi.org/10.1103/PhysRevD.111.074008}{{\em Phys. Rev. D} {\bfseries 111} no.~7, (2025) 074008}, \href{http://arxiv.org/abs/2406.04238}{{\ttfamily arXiv:2406.04238 [hep-ph]}}.

\bibitem{Caucal:2024vbv}
P.~Caucal, E.~Iancu, A.~H. Mueller, and F.~Yuan, ``{Jet Definition and Transverse-Momentum{\textendash}Dependent Factorization in Semi-inclusive Deep-Inelastic Scattering},'' \href{http://dx.doi.org/10.1103/PhysRevLett.134.061903}{{\em Phys. Rev. Lett.} {\bfseries 134} no.~6, (2025) 061903}, \href{http://arxiv.org/abs/2408.03129}{{\ttfamily arXiv:2408.03129 [hep-ph]}}.

\bibitem{Caucal:2024nsb}
P.~Caucal and F.~Salazar, ``{Dihadron correlations in small-$x$ DIS at NLO: transverse momentum dependent fragmentation},'' \href{http://arxiv.org/abs/2405.19404}{{\ttfamily arXiv:2405.19404 [hep-ph]}}.

\bibitem{Caucal:2025mth}
P.~Caucal, E.~Iancu, F.~Salazar, and F.~Yuan, ``{Gluon splitting at small $x$: a unified derivation for the JIMWLK, DGLAP and CSS equations},'' \href{http://arxiv.org/abs/2510.08454}{{\ttfamily arXiv:2510.08454 [hep-ph]}}.

\bibitem{Duan:2024nlr}
H.~Duan, A.~Kovner, and M.~Lublinsky, ``{Collins-Soper-Sterman Hamiltonian: High energy evolution of rapidity dependent observables},'' \href{http://dx.doi.org/10.1103/PhysRevD.111.054022}{{\em Phys. Rev. D} {\bfseries 111} no.~5, (2025) 054022}, \href{http://arxiv.org/abs/2407.15960}{{\ttfamily arXiv:2407.15960 [hep-ph]}}.

\bibitem{Duan:2024qev}
H.~Duan, A.~Kovner, and M.~Lublinsky, ``{Born-Oppenheimer renormalization group for high energy scattering: CSS, DGLAP and all that},'' \href{http://dx.doi.org/10.1007/JHEP08(2025)137}{{\em JHEP} {\bfseries 08} (2025) 137}, \href{http://arxiv.org/abs/2412.05097}{{\ttfamily arXiv:2412.05097 [hep-ph]}}.

\bibitem{Mukherjee:2023snp}
S.~Mukherjee, V.~V. Skokov, A.~Tarasov, and S.~Tiwari, ``{Unified description of DGLAP, CSS, and BFKL evolution: TMD factorization bridging large and small x},'' \href{http://dx.doi.org/10.1103/PhysRevD.109.034035}{{\em Phys. Rev. D} {\bfseries 109} no.~3, (2024) 034035}, \href{http://arxiv.org/abs/2311.16402}{{\ttfamily arXiv:2311.16402 [hep-ph]}}.

\bibitem{Stasto:2018rci}
A.~Stasto, S.-Y. Wei, B.-W. Xiao, and F.~Yuan, ``{On the Dihadron Angular Correlations in Forward $pA$ collisions},'' \href{http://dx.doi.org/10.1016/j.physletb.2018.08.011}{{\em Phys. Lett. B} {\bfseries 784} (2018) 301--306}, \href{http://arxiv.org/abs/1805.05712}{{\ttfamily arXiv:1805.05712 [hep-ph]}}.

\bibitem{Balitsky:2006wa}
I.~Balitsky, ``{Quark contribution to the small-x evolution of color dipole},'' \href{http://dx.doi.org/10.1103/PhysRevD.75.014001}{{\em Phys. Rev. D} {\bfseries 75} (2007) 014001}, \href{http://arxiv.org/abs/hep-ph/0609105}{{\ttfamily arXiv:hep-ph/0609105}}.

\bibitem{Kotko:2015ura}
P.~Kotko, K.~Kutak, C.~Marquet, E.~Petreska, S.~Sapeta, and A.~van Hameren, ``{Improved TMD factorization for forward dijet production in dilute-dense hadronic collisions},'' \href{http://dx.doi.org/10.1007/JHEP09(2015)106}{{\em JHEP} {\bfseries 09} (2015) 106}, \href{http://arxiv.org/abs/1503.03421}{{\ttfamily arXiv:1503.03421 [hep-ph]}}.

\bibitem{Altinoluk:2019fui}
T.~Altinoluk, R.~Boussarie, and P.~Kotko, ``{Interplay of the CGC and TMD frameworks to all orders in kinematic twist},'' \href{http://dx.doi.org/10.1007/JHEP05(2019)156}{{\em JHEP} {\bfseries 05} (2019) 156}, \href{http://arxiv.org/abs/1901.01175}{{\ttfamily arXiv:1901.01175 [hep-ph]}}.

\bibitem{Deganutti:2023qct}
F.~Deganutti, C.~Royon, and S.~Schlichting, ``{Forward dijet production at the LHC within an impact parameter dependent TMD approach},'' \href{http://dx.doi.org/10.1007/JHEP01(2024)159}{{\em JHEP} {\bfseries 01} (2024) 159}, \href{http://arxiv.org/abs/2311.01965}{{\ttfamily arXiv:2311.01965 [hep-ph]}}.

\bibitem{ALICE-PUBLIC-2019-005}
{\bfseries ALICE} Collaboration, ``{A Forward Calorimeter (FoCal) in the ALICE experiment},''. \url{http://cds.cern.ch/record/2696471}.

\bibitem{Jalilian-Marian:2004vhw}
J.~Jalilian-Marian and Y.~V. Kovchegov, ``{Inclusive two-gluon and valence quark-gluon production in DIS and pA},'' \href{http://dx.doi.org/10.1103/PhysRevD.71.079901}{{\em Phys. Rev. D} {\bfseries 70} (2004) 114017}, \href{http://arxiv.org/abs/hep-ph/0405266}{{\ttfamily arXiv:hep-ph/0405266}}. [Erratum: Phys.Rev.D 71, 079901 (2005)].

\bibitem{Baier:2005dv}
R.~Baier, A.~Kovner, M.~Nardi, and U.~A. Wiedemann, ``{Particle correlations in saturated QCD matter},'' \href{http://dx.doi.org/10.1103/PhysRevD.72.094013}{{\em Phys. Rev. D} {\bfseries 72} (2005) 094013}, \href{http://arxiv.org/abs/hep-ph/0506126}{{\ttfamily arXiv:hep-ph/0506126}}.

\bibitem{Dumitru:2005gt}
A.~Dumitru, A.~Hayashigaki, and J.~Jalilian-Marian, ``{The Color glass condensate and hadron production in the forward region},'' \href{http://dx.doi.org/10.1016/j.nuclphysa.2005.11.014}{{\em Nucl. Phys. A} {\bfseries 765} (2006) 464--482}, \href{http://arxiv.org/abs/hep-ph/0506308}{{\ttfamily arXiv:hep-ph/0506308}}.

\bibitem{Gelis:2003vh}
F.~Gelis and R.~Venugopalan, ``{Large mass q anti-q production from the color glass condensate},'' \href{http://dx.doi.org/10.1103/PhysRevD.69.014019}{{\em Phys. Rev. D} {\bfseries 69} (2004) 014019}, \href{http://arxiv.org/abs/hep-ph/0310090}{{\ttfamily arXiv:hep-ph/0310090}}.

\bibitem{Blaizot:2004wu}
J.~P. Blaizot, F.~Gelis, and R.~Venugopalan, ``{High-energy pA collisions in the color glass condensate approach. 1. Gluon production and the Cronin effect},'' \href{http://dx.doi.org/10.1016/j.nuclphysa.2004.07.005}{{\em Nucl. Phys. A} {\bfseries 743} (2004) 13--56}, \href{http://arxiv.org/abs/hep-ph/0402256}{{\ttfamily arXiv:hep-ph/0402256}}.

\bibitem{Blaizot:2004wv}
J.~P. Blaizot, F.~Gelis, and R.~Venugopalan, ``{High-energy pA collisions in the color glass condensate approach. 2. Quark production},'' \href{http://dx.doi.org/10.1016/j.nuclphysa.2004.07.006}{{\em Nucl. Phys. A} {\bfseries 743} (2004) 57--91}, \href{http://arxiv.org/abs/hep-ph/0402257}{{\ttfamily arXiv:hep-ph/0402257}}.

\bibitem{Cougoulic:2024jnd}
F.~Cougoulic, P.~Korcyl, and T.~Stebel, ``{Improving the solver for the Balitsky-Kovchegov evolution equation with Automatic Differentiation},'' \href{http://dx.doi.org/10.1016/j.cpc.2025.109616}{{\em Comput. Phys. Commun.} {\bfseries 313} (2025) 109616}, \href{http://arxiv.org/abs/2411.12739}{{\ttfamily arXiv:2411.12739 [hep-ph]}}.

\bibitem{Hou:2019qau}
T.-J. Hou {\em et al.}, ``{Progress in the CTEQ-TEA NNLO global QCD analysis},'' \href{http://arxiv.org/abs/1908.11394}{{\ttfamily arXiv:1908.11394 [hep-ph]}}.

\bibitem{Sato:2019yez}
{\bfseries JAM} Collaboration, N.~Sato, C.~Andres, J.~J. Ethier, and W.~Melnitchouk, ``{Strange quark suppression from a simultaneous Monte Carlo analysis of parton distributions and fragmentation functions},'' \href{http://dx.doi.org/10.1103/PhysRevD.101.074020}{{\em Phys. Rev. D} {\bfseries 101} no.~7, (2020) 074020}, \href{http://arxiv.org/abs/1905.03788}{{\ttfamily arXiv:1905.03788 [hep-ph]}}.

\bibitem{Gao:2024nkz}
J.~Gao, C.~Liu, X.~Shen, H.~Xing, and Y.~Zhao, ``{Simultaneous Determination of Fragmentation Functions and Test on Momentum Sum Rule},'' \href{http://dx.doi.org/10.1103/PhysRevLett.132.261903}{{\em Phys. Rev. Lett.} {\bfseries 132} no.~26, (2024) 261903}, \href{http://arxiv.org/abs/2401.02781}{{\ttfamily arXiv:2401.02781 [hep-ph]}}.

\bibitem{deFlorian:2007aj}
D.~de~Florian, R.~Sassot, and M.~Stratmann, ``{Global analysis of fragmentation functions for pions and kaons and their uncertainties},'' \href{http://dx.doi.org/10.1103/PhysRevD.75.114010}{{\em Phys. Rev. D} {\bfseries 75} (2007) 114010}, \href{http://arxiv.org/abs/hep-ph/0703242}{{\ttfamily arXiv:hep-ph/0703242}}.

\bibitem{Albacete:2010sy}
J.~L. Albacete, N.~Armesto, J.~G. Milhano, P.~Quiroga-Arias, and C.~A. Salgado, ``{AAMQS: A non-linear QCD analysis of new HERA data at small-x including heavy quarks},'' \href{http://dx.doi.org/10.1140/epjc/s10052-011-1705-3}{{\em Eur. Phys. J. C} {\bfseries 71} (2011) 1705}, \href{http://arxiv.org/abs/1012.4408}{{\ttfamily arXiv:1012.4408 [hep-ph]}}.

\bibitem{Lappi:2013zma}
T.~Lappi and H.~M\"antysaari, ``{Single inclusive particle production at high energy from HERA data to proton-nucleus collisions},'' \href{http://dx.doi.org/10.1103/PhysRevD.88.114020}{{\em Phys. Rev. D} {\bfseries 88} (2013) 114020}, \href{http://arxiv.org/abs/1309.6963}{{\ttfamily arXiv:1309.6963 [hep-ph]}}.

\bibitem{Gelis:2006tb}
F.~Gelis, A.~M. Stasto, and R.~Venugopalan, ``{Limiting fragmentation in hadron-hadron collisions at high energies},'' \href{http://dx.doi.org/10.1140/epjc/s10052-006-0020-x}{{\em Eur. Phys. J. C} {\bfseries 48} (2006) 489--500}, \href{http://arxiv.org/abs/hep-ph/0605087}{{\ttfamily arXiv:hep-ph/0605087}}.

\bibitem{Kowalski:2007rw}
H.~Kowalski, T.~Lappi, and R.~Venugopalan, ``{Nuclear enhancement of universal dynamics of high parton densities},'' \href{http://dx.doi.org/10.1103/PhysRevLett.100.022303}{{\em Phys. Rev. Lett.} {\bfseries 100} (2008) 022303}, \href{http://arxiv.org/abs/0705.3047}{{\ttfamily arXiv:0705.3047 [hep-ph]}}.

\bibitem{Sun:2014dqm}
P.~Sun, J.~Isaacson, C.~P. Yuan, and F.~Yuan, ``{Nonperturbative functions for SIDIS and Drell{\textendash}Yan processes},'' \href{http://dx.doi.org/10.1142/S0217751X18410063}{{\em Int. J. Mod. Phys. A} {\bfseries 33} no.~11, (2018) 1841006}, \href{http://arxiv.org/abs/1406.3073}{{\ttfamily arXiv:1406.3073 [hep-ph]}}.

\bibitem{Prokudin:2015ysa}
A.~Prokudin, P.~Sun, and F.~Yuan, ``{Scheme dependence and transverse momentum distribution interpretation of Collins{\textendash}Soper{\textendash}Sterman resummation},'' \href{http://dx.doi.org/10.1016/j.physletb.2015.09.064}{{\em Phys. Lett. B} {\bfseries 750} (2015) 533--538}, \href{http://arxiv.org/abs/1505.05588}{{\ttfamily arXiv:1505.05588 [hep-ph]}}.

\bibitem{Echevarria:2020hpy}
M.~G. Echevarria, Z.-B. Kang, and J.~Terry, ``{Global analysis of the Sivers functions at NLO+NNLL in QCD},'' \href{http://dx.doi.org/10.1007/JHEP01(2021)126}{{\em JHEP} {\bfseries 01} (2021) 126}, \href{http://arxiv.org/abs/2009.10710}{{\ttfamily arXiv:2009.10710 [hep-ph]}}.

\bibitem{Kang:2025vjk}
Z.-B. Kang, R.~Kao, M.~Li, and J.~Penttala, ``{Transverse energy-energy correlators at small x for photon-hadron production},'' \href{http://dx.doi.org/10.1103/vllq-fhr7}{{\em Phys. Rev. D} {\bfseries 112} no.~7, (2025) 076006}, \href{http://arxiv.org/abs/2504.00069}{{\ttfamily arXiv:2504.00069 [hep-ph]}}.

\bibitem{Ganguli:2025aqa}
I.~Ganguli and P.~Kotko, ``{Transverse energy-energy and azimuthal correlations in forward {\ensuremath{\gamma}}-hadron production in proton-proton and proton-lead collisions at the LHC},'' \href{http://dx.doi.org/10.1103/pktk-k5g1}{{\em Phys. Rev. D} {\bfseries 112} no.~11, (2025) 114020}, \href{http://arxiv.org/abs/2507.23435}{{\ttfamily arXiv:2507.23435 [hep-ph]}}.

\bibitem{Moult:2025nhu}
I.~Moult and H.~X. Zhu, ``{Energy Correlators: A Journey From Theory to Experiment},'' \href{http://arxiv.org/abs/2506.09119}{{\ttfamily arXiv:2506.09119 [hep-ph]}}.

\bibitem{Marquet:2017xwy}
C.~Marquet, C.~Roiesnel, and P.~Taels, ``{Linearly polarized small-$x$ gluons in forward heavy-quark pair production},'' \href{http://dx.doi.org/10.1103/PhysRevD.97.014004}{{\em Phys. Rev. D} {\bfseries 97} no.~1, (2018) 014004}, \href{http://arxiv.org/abs/1710.05698}{{\ttfamily arXiv:1710.05698 [hep-ph]}}.

\bibitem{Altinoluk:2021ygv}
T.~Altinoluk, C.~Marquet, and P.~Taels, ``{Low-x improved TMD approach to the lepto- and hadroproduction of a heavy-quark pair},'' \href{http://dx.doi.org/10.1007/JHEP06(2021)085}{{\em JHEP} {\bfseries 06} (2021) 085}, \href{http://arxiv.org/abs/2103.14495}{{\ttfamily arXiv:2103.14495 [hep-ph]}}.

\bibitem{Marquet:2025jdr}
C.~Marquet, Y.~Shi, and B.-W. Xiao, ``{Unified Resummation of Soft Gluon Radiation in Heavy Meson Pair Photoproduction},'' \href{http://arxiv.org/abs/2510.18949}{{\ttfamily arXiv:2510.18949 [hep-ph]}}.

\bibitem{Giacalone:2018fbc}
G.~Giacalone and C.~Marquet, ``{Signature of gluon saturation in forward di-hadron correlations at the Large Hadron Collider},'' \href{http://dx.doi.org/10.1016/j.nuclphysa.2018.10.009}{{\em Nucl. Phys. A} {\bfseries 982} (2019) 291--294}, \href{http://arxiv.org/abs/1807.06388}{{\ttfamily arXiv:1807.06388 [hep-ph]}}.

\bibitem{ALICE:2012eyl}
{\bfseries ALICE} Collaboration, B.~Abelev {\em et al.}, ``{Long-range angular correlations on the near and away side in $p$-Pb collisions at $\sqrt{s_{NN}}=5.02$ TeV},'' \href{http://dx.doi.org/10.1016/j.physletb.2013.01.012}{{\em Phys. Lett. B} {\bfseries 719} (2013) 29--41}, \href{http://arxiv.org/abs/1212.2001}{{\ttfamily arXiv:1212.2001 [nucl-ex]}}.

\bibitem{LHCb:2015coe}
{\bfseries LHCb} Collaboration, R.~Aaij {\em et al.}, ``{Measurements of long-range near-side angular correlations in $\sqrt{s_{\text{NN}}}=5$TeV proton-lead collisions in the forward region},'' \href{http://dx.doi.org/10.1016/j.physletb.2016.09.064}{{\em Phys. Lett. B} {\bfseries 762} (2016) 473--483}, \href{http://arxiv.org/abs/1512.00439}{{\ttfamily arXiv:1512.00439 [nucl-ex]}}.

\bibitem{Iancu:2022gpw}
E.~Iancu and Y.~Mulian, ``{Dihadron production in DIS at NLO: the real corrections},'' \href{http://dx.doi.org/10.1007/JHEP07(2023)121}{{\em JHEP} {\bfseries 07} (2023) 121}, \href{http://arxiv.org/abs/2211.04837}{{\ttfamily arXiv:2211.04837 [hep-ph]}}.

\bibitem{Caucal:2025xxh}
P.~Caucal, M.~G. Morales, E.~Iancu, F.~Salazar, and F.~Yuan, ``{Unveiling the sea: universality of the transverse momentum dependent quark distributions at small $x$},'' \href{http://arxiv.org/abs/2503.16162}{{\ttfamily arXiv:2503.16162 [hep-ph]}}.

\bibitem{Chirilli:2011km}
G.~A. Chirilli, B.-W. Xiao, and F.~Yuan, ``{One-loop Factorization for Inclusive Hadron Production in $pA$ Collisions in the Saturation Formalism},'' \href{http://dx.doi.org/10.1103/PhysRevLett.108.122301}{{\em Phys. Rev. Lett.} {\bfseries 108} (2012) 122301}, \href{http://arxiv.org/abs/1112.1061}{{\ttfamily arXiv:1112.1061 [hep-ph]}}.

\bibitem{Shi:2021hwx}
Y.~Shi, L.~Wang, S.-Y. Wei, and B.-W. Xiao, ``{Pursuing the Precision Study for Color Glass Condensate in Forward Hadron Productions},'' \href{http://dx.doi.org/10.1103/PhysRevLett.128.202302}{{\em Phys. Rev. Lett.} {\bfseries 128} no.~20, (2022) 202302}, \href{http://arxiv.org/abs/2112.06975}{{\ttfamily arXiv:2112.06975 [hep-ph]}}.

\bibitem{Wang:2022zdu}
L.~Wang, L.~Chen, Z.~Gao, Y.~Shi, S.-Y. Wei, and B.-W. Xiao, ``{Forward inclusive jet productions in pA collisions},'' \href{http://dx.doi.org/10.1103/PhysRevD.107.016016}{{\em Phys. Rev. D} {\bfseries 107} no.~1, (2023) 016016}, \href{http://arxiv.org/abs/2211.08322}{{\ttfamily arXiv:2211.08322 [hep-ph]}}.

\bibitem{Kang:2011bp}
Z.-B. Kang, I.~Vitev, and H.~Xing, ``{Dihadron momentum imbalance and correlations in d+Au collisions},'' \href{http://dx.doi.org/10.1103/PhysRevD.85.054024}{{\em Phys. Rev. D} {\bfseries 85} (2012) 054024}, \href{http://arxiv.org/abs/1112.6021}{{\ttfamily arXiv:1112.6021 [hep-ph]}}.

\bibitem{Fu:2023jqv}
Y.~Fu, Z.-B. Kang, F.~Salazar, X.-N. Wang, and H.~Xing, ``{Correspondence between Color Glass Condensate and High-Twist Formalism},'' \href{http://dx.doi.org/10.1103/PhysRevLett.135.032301}{{\em Phys. Rev. Lett.} {\bfseries 135} no.~3, (2025) 032301}, \href{http://arxiv.org/abs/2310.12847}{{\ttfamily arXiv:2310.12847 [hep-ph]}}.

\bibitem{Fu:2024sba}
Y.~Fu, Z.-B. Kang, F.~Salazar, X.-N. Wang, and H.~Xing, ``{Color glass condensate meets high twist expansion},'' \href{http://dx.doi.org/10.1103/ckhv-5213}{{\em Phys. Rev. D} {\bfseries 112} no.~1, (2025) 014029}, \href{http://arxiv.org/abs/2406.01684}{{\ttfamily arXiv:2406.01684 [hep-ph]}}.

\bibitem{Arleo:2025oos}
F.~Arleo {\em et al.}, ``{Nuclear Cold QCD: Review and Future Strategy},'' \href{http://arxiv.org/abs/2506.17454}{{\ttfamily arXiv:2506.17454 [hep-ph]}}.

\bibitem{Iancu:2011ns}
E.~Iancu and D.~N. Triantafyllopoulos, ``{Higher-point correlations from the JIMWLK evolution},'' \href{http://dx.doi.org/10.1007/JHEP11(2011)105}{{\em JHEP} {\bfseries 11} (2011) 105}, \href{http://arxiv.org/abs/1109.0302}{{\ttfamily arXiv:1109.0302 [hep-ph]}}.

\end{thebibliography}\endgroup

\appendix

\begin{widetext}

\section{Small-x gluon TMDs in the Gaussian approximation}

\label{app:TMDs-Gaussian}

In the Gaussian approximation, the correlator of Wilson lines and their derivatives can be expressed in terms of the two-point function of light-like Wilson lines in the fundamental representation \cite{Dominguez:2011wm,Iancu:2011ns}. By assuming translational invariance for this two-point function, so that it depends only on the relative transverse separation rather than on the individual coordinates, and employing the mean-field approximation, we are able to simplify the expression for the unpolarized gluon distributions defined in Eq.~(\ref{eq:tmds}). 

The TMDs can then be obtained from the following expressions:
\begin{align}
    \alpha_s \Fcal_{qg}^{(1)}(\kt,x) &= \frac{N_c S_\perp }{2\pi^2} \int \frac{r_\perp \der r_\perp}{2\pi} J_0(k_\perp r_\perp) \nabla^2_\perp \left[ 1-\langle S^{(2)}_F(r_\perp)\rangle_{x}\right] \label{eq:Fqg1}, \\
    \alpha_s \Fcal_{qg}^{(2)}(\kt,x) &= \frac{C_F S_\perp }{2\pi^2} \int \frac{r_\perp \der r_\perp}{2\pi} J_0(k_\perp r_\perp) \Kcal(x,r_\perp)\left[1 - \left( \langle S^{(2)}_F(r_\perp)\rangle_{x} \right)^{N_c/C_F} \right]\langle S^{(2)}_F(r_\perp)\rangle_{x} \label{eq:Fqg2},\\
    \alpha_s \Fcal_{gg}^{(1)}(\kt,x) &= \frac{N_c S_\perp }{2\pi^2} \int \frac{r_\perp \der r_\perp}{2\pi} J_0(k_\perp r_\perp) \langle S^{(2)}_F(r_\perp)\rangle_{x} \nabla^2_\perp \left[1 - \langle S^{(2)}_F(r_\perp)\rangle_{x} \right] \label{eq:Fgg1},  \\
    \alpha_s \Fcal_{gg}^{(2)}(\kt,x) & = \alpha_s \Fcal_{gg}^{(1)}(\kt,x) - \alpha_s \Fcal_{adj}(\kt,x),\label{eq:Fgg2} \\
    \alpha_s \Fcal_{gg}^{(3)}(\kt,x) &= \frac{C_F S_\perp }{2\pi^2} \int \frac{r_\perp \der r_\perp}{2\pi} J_0(k_\perp r_\perp) \Kcal(x,r_\perp)\left[1 - \left( \langle S^{(2)}_F(r_\perp)\rangle_{x} \right)^{N_c/C_F} \right]\left(\langle S^{(2)}_F(r_\perp)\rangle_{x}\right)^{2} \label{eq:Fgg3},\\
    \alpha_s \Fcal_{Adj}(\kt,x) &= \frac{C_F S_\perp }{2\pi^2} \int \frac{r_\perp \der r_\perp}{2\pi} J_0(k_\perp r_\perp) \nabla^2_\perp \left[ 1-\left(\langle S^{(2)}_F(r_\perp)\rangle_{x}\right)^{N_c/C_F}\right]. 
     \label{eq:FWW}
\end{align}
Here $S_\perp$ is the overall area factor, $C_F = (N_c^2 -1)/(2 N_c)$, and $N_c=3$, $J_0(\zeta)$ is the Bessel function of the first kind of order zero, $\nabla^2_\perp = \frac{\partial^2}{\partial r_\perp^2}  + \frac{1}{r_\perp} \frac{\partial}{\partial r_\perp}$, $ \Kcal(x,r_\perp)= \frac{\nabla^2_\perp \Gamma(x,r_\perp)}{\Gamma(x,r_\perp)}$, and $\Gamma(x,r_\perp) = -\ln\left[\langle S^{(2)}_F(r_\perp)\rangle_{x}\right] $. It is important to note that the mean-field approximation neglects sub-leading contributions in $1/N_c$. For consistency, we work in the large‑$N_c$ limit, which entails replacing the fundamental Casimir, $C_F$, by $N_c/2$.

\section{Numerical results for small-$x$ TMDs}

In this appendix we present the numerical results for the five small-$x$ TMDs in Eq.\,\eqref{eq:tmds} needed to compute dihadron correlations in p-p and p-$A$ collisions. We compute them following the Gaussian approximation in Appendix \ref{app:TMDs-Gaussian} and using the solution to running coupling Balitsky-Kovchegov equations (for details see Sec.\,\ref{sec:np-input}). As we assume translational invariance of the dipole, we present results for the TMDs, normalized by the overall area factor $S_\perp$.

Figure $\ref{fig:TMDs}$ shows these five TMDs for both the proton and Lead at two values of $x_g$. As it is well known, the various TMDs differ significantly at low $k_T$ but converge to the same perturbative tail in the high-$k_T$ (or “dilute”) region. At low $k_T$, the different behavior arises from the distinct gauge link structures used to define each TMD. The transition between low and high $k_{T}$ regimes is dictated by the saturation scale. This is evident when comparing TMDs at the value of $x$ for proton and Lead. Additionally, as the TMDs are evolved to smaller $x_g$, they shift towards higher $k_T$, reflecting the fact that the saturation scale increases after small‑$x$ evolution.
\begin{figure}[H]
    \centering
    \includegraphics[width=0.45\columnwidth]{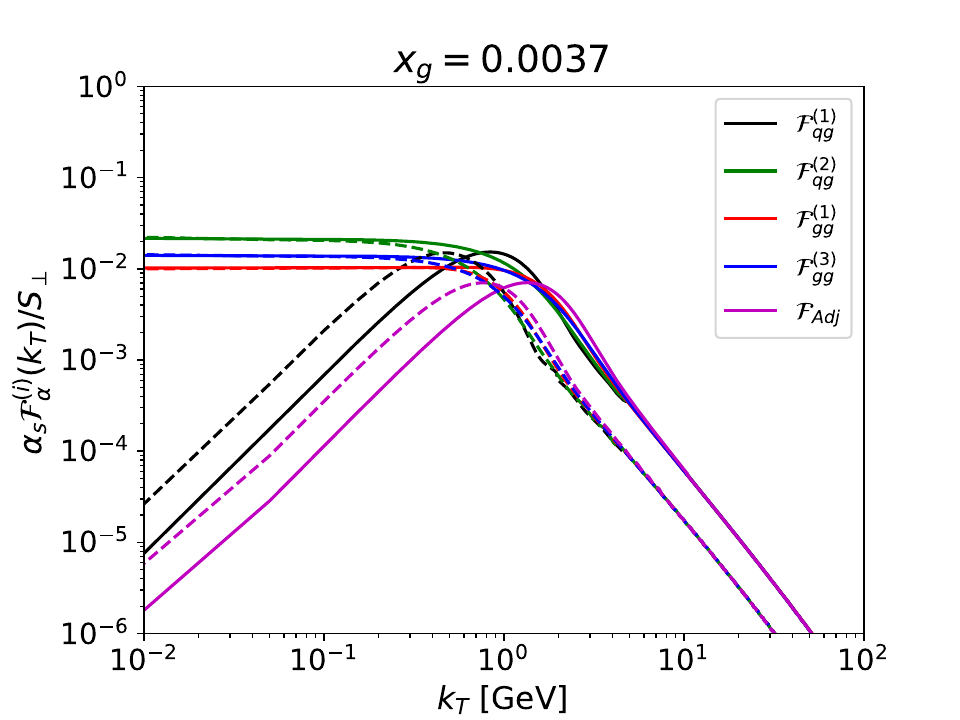}
    \includegraphics[width=0.45\columnwidth]{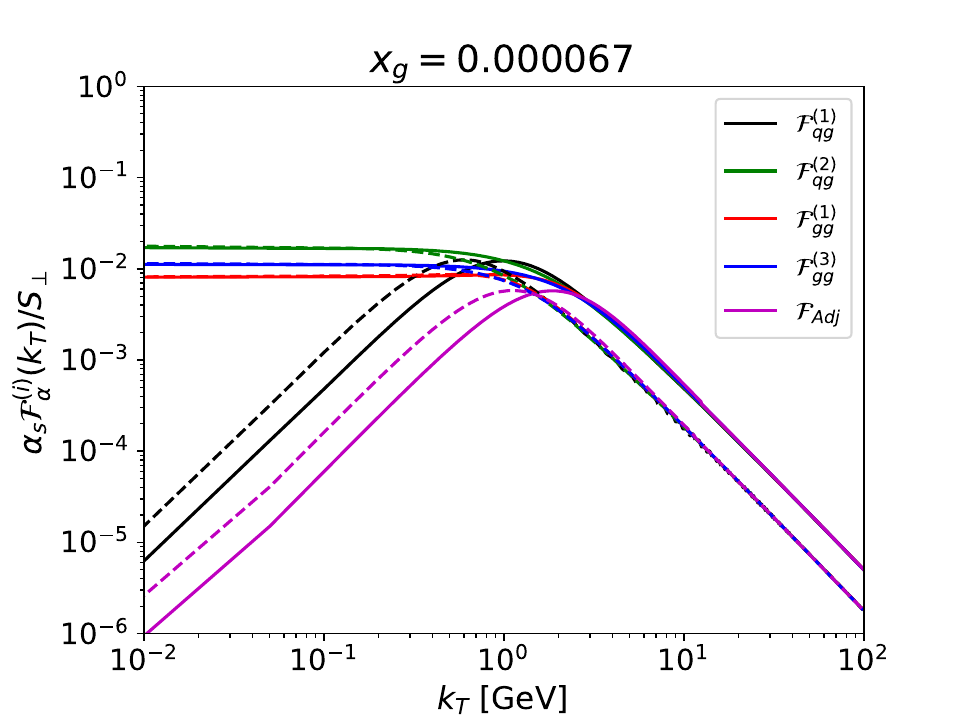}
    \caption{The five small-$x$ gluon TMDs for the proton (dashed lines) and Lead (solid lines) at gluon $x_g = 3.7\times 10^{-3}$ (left) and $x_g=6.7\times 10^{-5}$ (right) }
    \label{fig:TMDs}
\end{figure}

\section{Partonic production cross-sections in the ITMD}
\label{app:ITMD-hard-factor}

In the improved TMD framework,  hard factors are computed from matrix elements where the small-$x$ gluon is off-shell \cite{Kotko:2015ura}. The explicit expressions for the improved TMD hard factors that appear in Eq. (\ref{eq:double-inclusive}) are: 
\begin{align}
    \Hcal^{(1)}_{qg \to qg} &= \frac{\alpha_s^2(1-z) \left[1 + (1-z)^2 \right] }{2\Pt^2 }\left[ \frac{(1-z)^2 }{\pqt^2} + \frac{1}{N_c^2} \frac{1}{\pgt^2} \left( \frac{\Pt^2}{\pqt^2} - z^2 \right) \right] \,, \\
    \Hcal^{(2)}_{qg \to qg} &= \frac{\alpha_s^2 C_F}{N_c}\frac{(1-z) \left[1 + (1-z)^2 \right] }{\pgt^2 \pqt ^2}, \\
    \Hcal^{(1)}_{gg \to q\bar{q}} &= \frac{\alpha_s^2}{2N_c}\frac{z(1-z)\left[z^2 + (1-z)^2 \right]}{\Pt^2} \left[ \frac{(1-z)^2}{\pqbt^2} + \frac{z^2}{\pqt^2} \right] \,, \\
    \Hcal^{(2)}_{gg \to q\bar{q}} &= \frac{\alpha_s^2}{2N_c} \frac{z(1-z)\left[z^2 + (1-z)^2 \right]}{\Pt^2} \left[\frac{\Pt^2}{\pqt^2 \pqbt^2} - \frac{(1-z)^2}{\pqbt^2} - \frac{z^2}{\pqt^2} \right] \,, \\
    \Hcal^{(1)}_{gg \to gg} &= \frac{\alpha_s^2 N_c}{2C_F}\frac{2 z(1-z) }{\Pt^2} \left[\frac{z}{1-z} + \frac{1-z}{z} + z(1-z) \right]  \left[ \frac{(1-z)^2}{\pgttwo^2} + \frac{z^2}{\pgtone^2} \right] \,, \\
    \Hcal^{(2)}_{gg \to gg} & = \frac{\alpha_s^2 N_c}{2C_F}\frac{2 z(1-z) }{\Pt^2} \left[\frac{z}{1-z} + \frac{1-z}{z} + z(1-z) \right]  \left[\frac{\Pt^2}{\pgtone^2 \pgttwo^2} - \frac{(1-z)^2}{\pgttwo^2} - \frac{z^2}{\pgtone^2} \right] \,.
\end{align}
Here, $\boldsymbol{k_{i}}$ denotes the transverse momentum of the corresponding partons. In the numerical implementation, one power of $\alpha_s$ in the hard factors is canceled by the $1/\alpha_s$ prefactor multiplying the gluon TMDs, whereas the remaining strong coupling is evaluated at the hard scale $Q$.

\section{Supplemental plots}
\label{app:plots}
\begin{figure}
    \centering    \includegraphics[width=0.45\columnwidth]{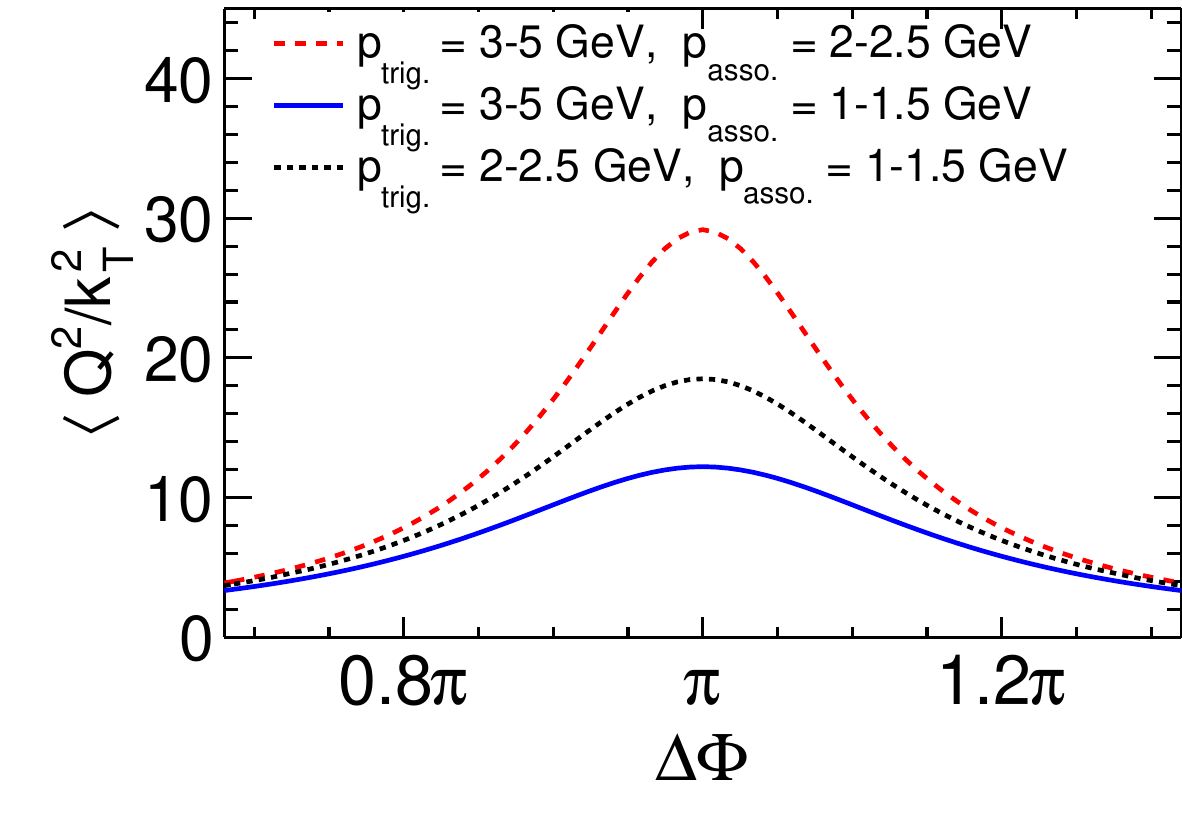}
    \caption{ The average ratio $\langle Q^2/k_T^2\rangle$ as a function of $\Delta\Phi$ for different bins, with $Q^2 = x_p x_g s$ chosen as the scale entering the upper limit of the Sudakov integration.}
    \label{fig:ratio_Q2_kT}
\end{figure}

\begin{figure}[H]
    \centering
    \includegraphics[width=0.37\columnwidth]{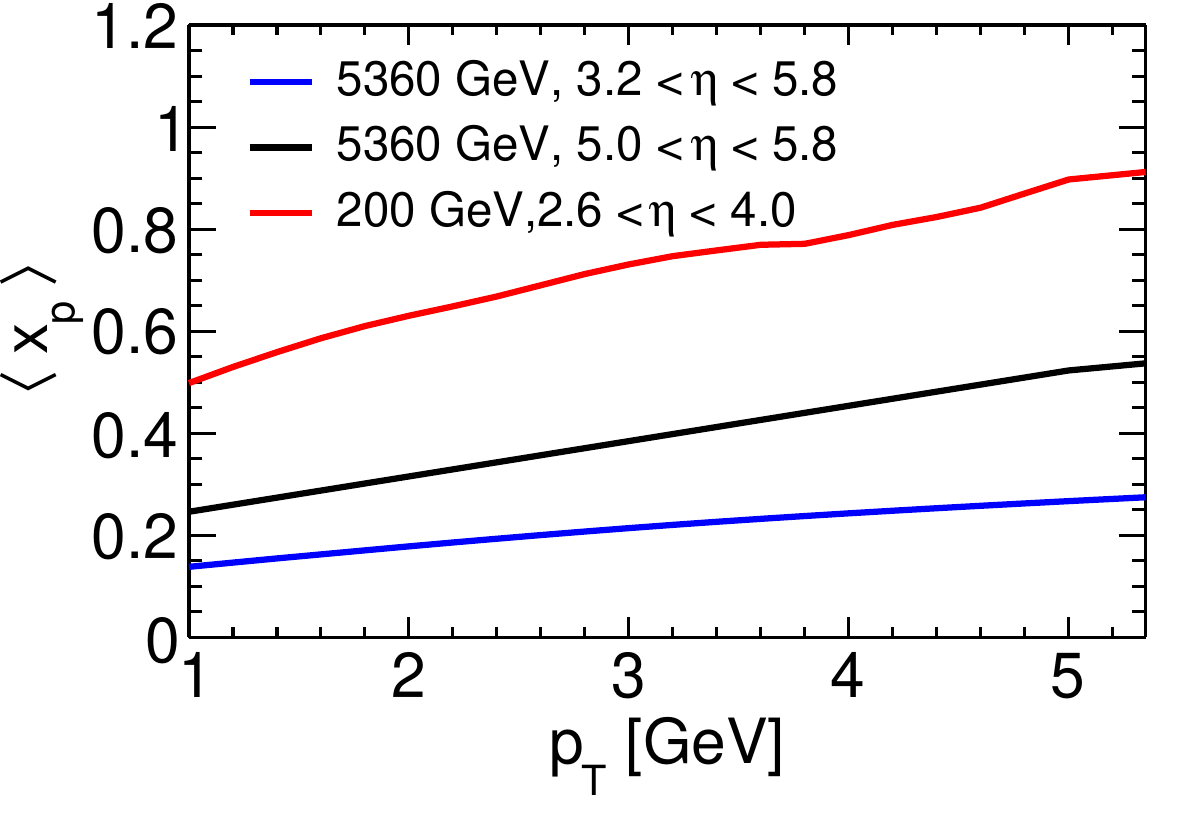}
    \includegraphics[width=0.37\columnwidth]{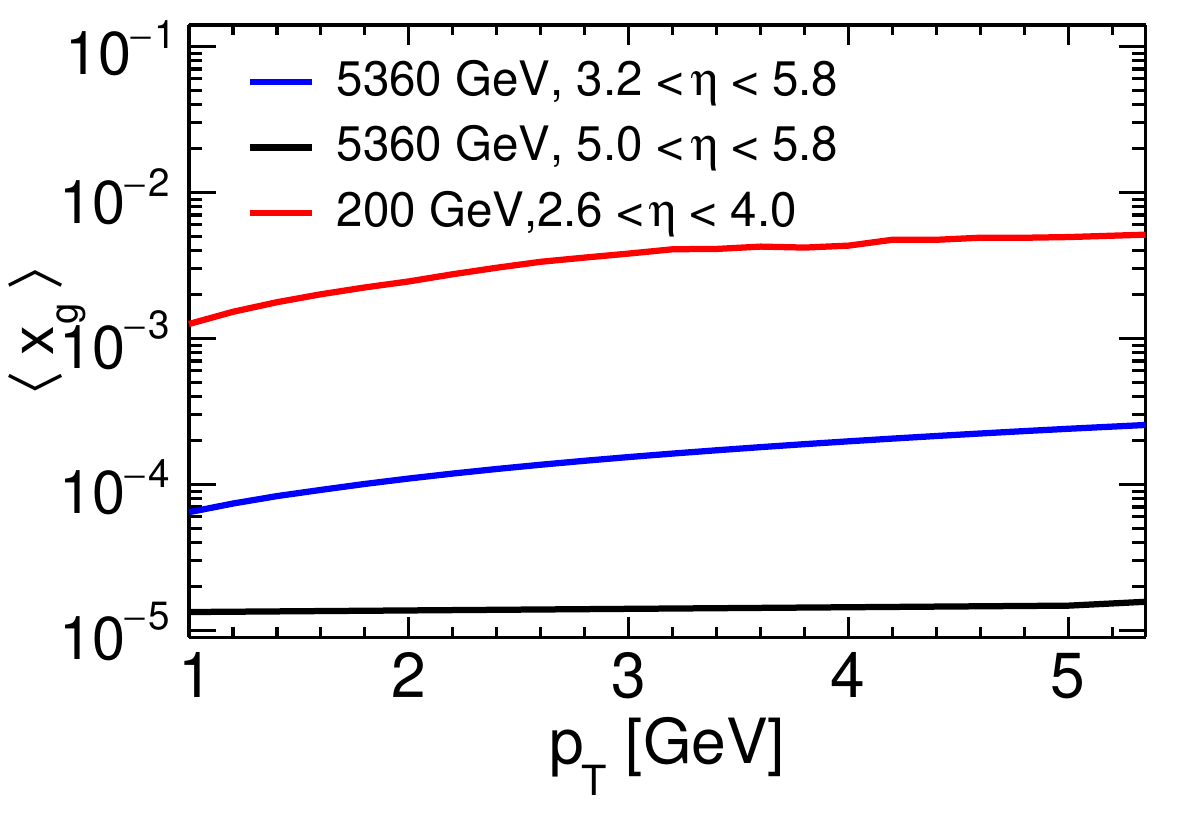}
    \includegraphics[width=0.37\columnwidth]{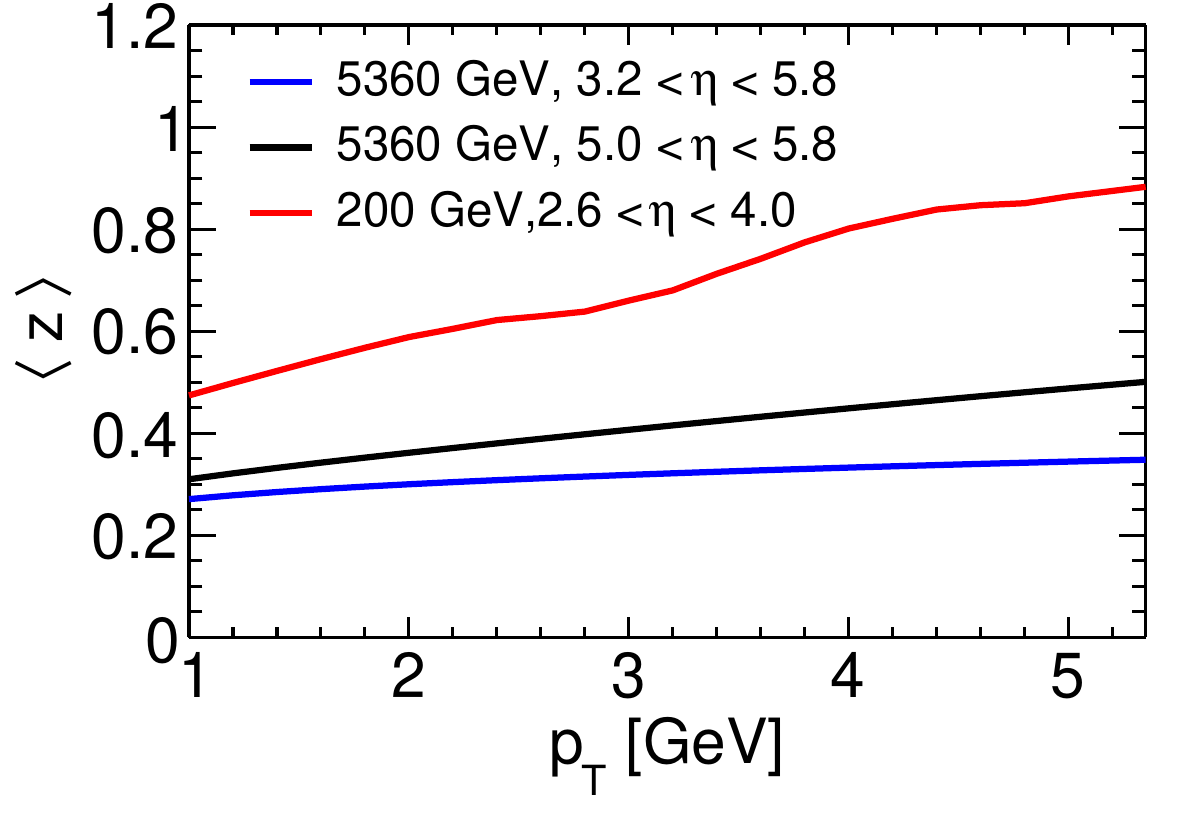}
    \caption{The averaged $\langle x_p\rangle$ (left) of the proton,    averaged gluon $\langle x_g\rangle$  (right) of the nucleus, and fragmentation energy fraction of hadron $\langle z\rangle$ (lower) at $p_T = p_{T1} = p_{T2}$, for the STAR (p-Au) and FoCal (p-Pb) kinematics.}
    \label{fig:xp}
\end{figure}

\begin{figure}[H]
    \centering
    \includegraphics[width=0.37\columnwidth]{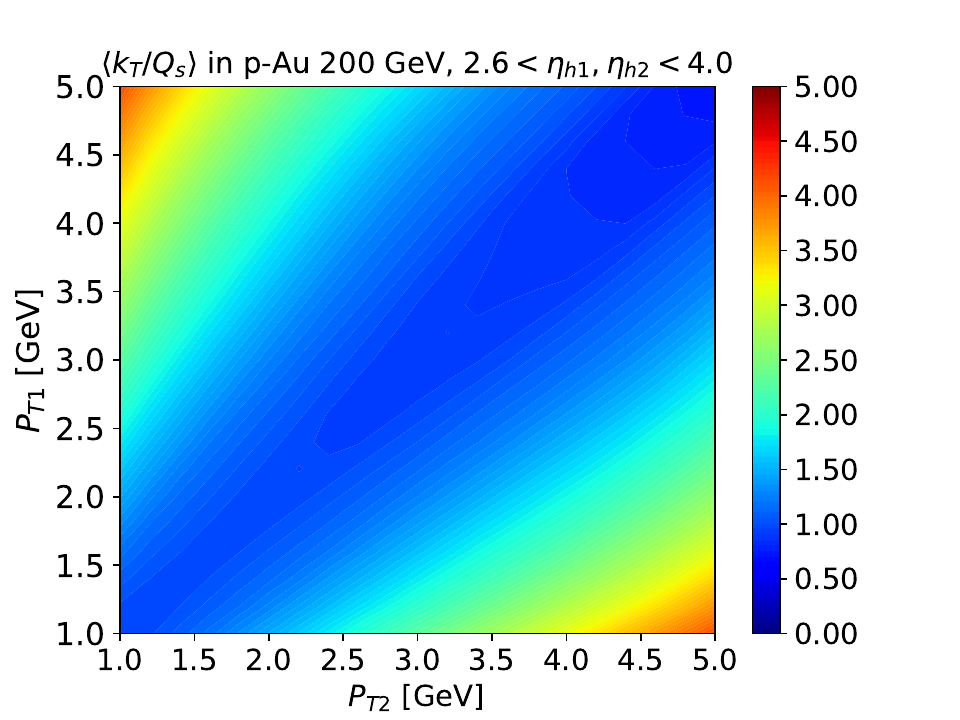}
    \includegraphics[width=0.37\columnwidth]{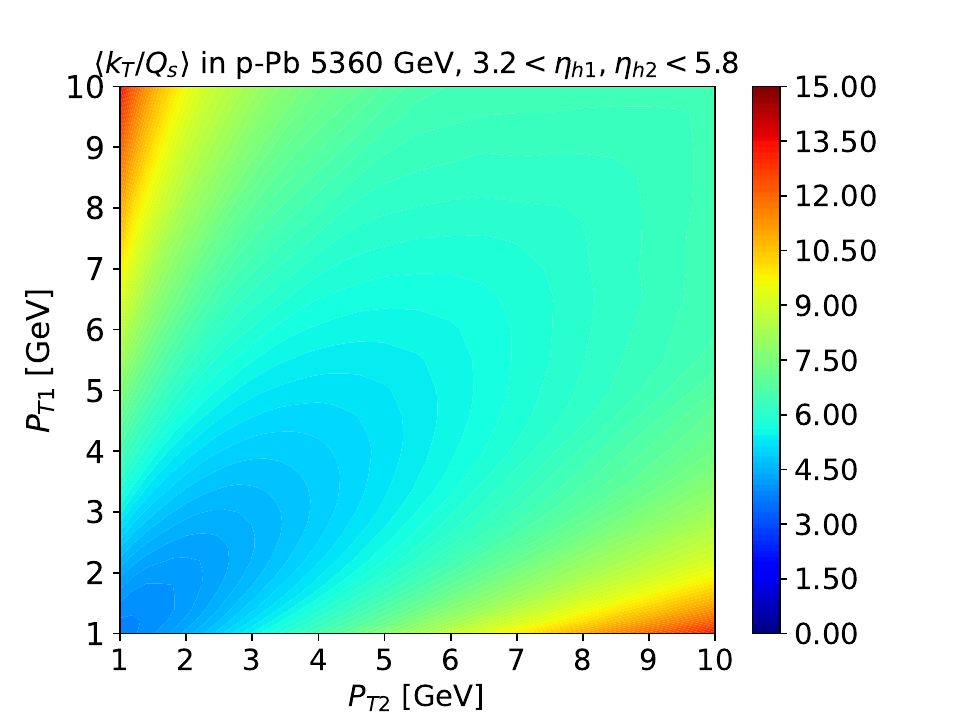}
    \includegraphics[width=0.37\columnwidth]{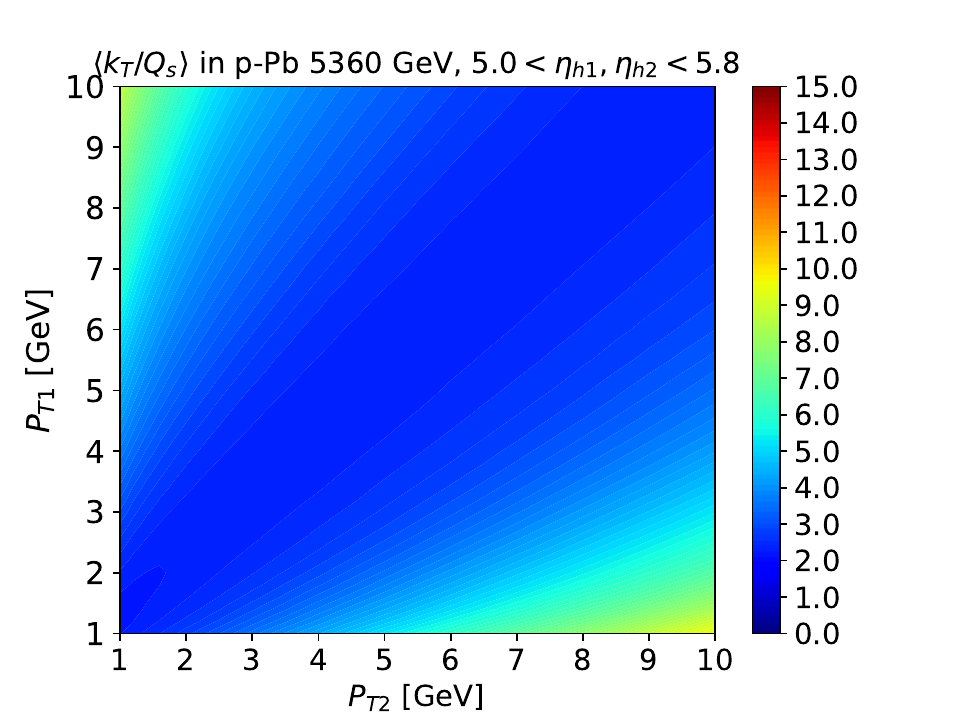}
    \caption{The averaged values of the ratio $k_T$ to $Q_s$, $\langle k_T/Q_s\rangle$,  at the STAR and FoCal kinematics.  }
    \label{fig:kT_Qs}
\end{figure}

\end{widetext}

\end{document}